\documentclass[aps,prx,showpacs,notitlepage,twocolumn, floatfix,superscriptaddress,nofootinbib]{revtex4-2}
\usepackage[utf8]{inputenc}
\usepackage{amssymb}
\usepackage{amsmath}
\usepackage{braket}
\usepackage{amsfonts}
\usepackage{graphicx}
\usepackage[colorlinks,linkcolor=blue,anchorcolor=blue,citecolor=blue,urlcolor=blue]{hyperref}
\usepackage[titletoc]{appendix}
\graphicspath{{Figures/}}

\def\bea{\begin{equation}\begin{aligned}}
\def\eea{\end{aligned}\end{equation}}
\def\bn{\bar{n}}
\def\F{\mathcal{F}}
\def\S{\mathcal{S}}
\def\H{\mathbb{H}}

\def\Q{\mathcal{Q}}

\def\o{\text{out}}
\def\i{\text{in}}
\def\w{\vec{w}}
\def\n{\vec{n}}

\def\tr{\text{tr}}

\begin{document}

\title{Charge transport, information scrambling and quantum operator-coherence in a many-body system with U(1) symmetry}

\author{Lakshya Agarwal}
\affiliation{Department of Physics \& Astronomy, Texas A\&M University, College Station, Texas 77843, USA}

\author{Subhayan Sahu}
\affiliation{Condensed Matter Theory Center and Department of Physics, University of Maryland, College Park, MD 20742, USA}
\affiliation{Perimeter Institute for Theoretical Physics, Waterloo, Ontario N2L 2Y5, Canada}

\author{Shenglong Xu}
\affiliation{Department of Physics \& Astronomy, Texas A\&M University, College Station, Texas 77843, USA}

\begin{abstract}

In this work, we derive an exact hydrodynamical description for the coupled, charge and operator dynamics, in a quantum many-body system with U(1) symmetry. 
Using an emergent symmetry in the complex Brownian SYK model with charge conservation, we map the operator dynamics in the model to the imaginary-time dynamics of an SU(4) spin-chain.
 We utilize the emergent SU(4) description to demonstrate that the U(1) symmetry causes quantum-coherence to persist even after disorder-averaging, in sharp contrast to models without symmetries.
In line with this property, we write down a `restricted’ Fokker-Planck equation for the out-of-time ordered correlator (OTOC) in the large-$N$ limit, which permits a classical probability description strictly in the incoherent sector of the global operator-space. We then exploit this feature to describe the OTOC in terms of a Fisher-Kolmogorov-Petrovsky-Piskun (FKPP)-equation which couples the operator with the charge and is valid at all time-scales and for arbitrary charge-density profiles. The coupled equations obtained belong to a class of models also used to describe the population dynamics of bacteria embedded in a diffusive media. We simulate them to explore operator-dynamics in a background of non-uniform charge configuration, which reveals that the charge transport can strongly affect dynamics of operators, including those that have no overlap with the charge.

\end{abstract}
\maketitle

\tableofcontents

\section{Introduction}


In many-body dynamics, a fine-grained approach to study thermalization~\cite{Deutsch1991,Srednicki1994,rigol2008thermalization, Polkovnikov2011} comes from the study of operator-dynamics~\cite{Nahum_2018, von2018operator,parker2019universal}. This is also related to the concept of information scrambling, where local information flows to non-local degrees of freedom and is not retrievable via local measurements~\cite{Hosur_2016, keselman2021scrambling, knap2018entanglement, Sekino_2008, Hayden_2007, shenker2015stringy}. The notions of operator growth and information scrambling can be precisely measured using the out-of-time-ordered correlator (OTOC):
\bea
\label{Eq:OTOC_definition}
\F(W(t),V) = \frac{1}{\tr(I)} \tr(W^{\dagger}(t) V^{\dagger} W(t) V)
\eea
This quantity is often used to measure the size of the evolving operator $W(t)$ and shows interesting features such as Lyapunov growth and butterfly velocity~\cite{Maldacena_2016, Sekino_2008, shenker2014black, Lashkari_2013, Polchinski2016,Maldacena_2016_SYK, Gross_2017, Gu_2020, han2019quantum, shenker2014black, roberts2016lieb, aleiner2016microscopic, luitz2017information, khemani2018velocity, Xu_2019_nature, lin2018out}.
The OTOC is experimentally accessible on various platforms~\cite{Li_2017, wei2018exploring, nie2019detecting, sanchez2020perturbation,Geller_2018,braumuller2022probing, mi2021information,Blok_2021,Garttner_2017,Landsman_2019,joshi2020quantum}, where it is often used as a measure of information scrambling.


Previous work on random circuits and noisy-driven models has established several related effective classical models of operator dynamics, including biased random walk~\cite{Nahum_2018}, reaction diffusion process~\cite{Xu_2019} and population dynamics~\cite{qi2019quantum}. These classical models are used to fit and understand experimental data of the OTOC on quantum platforms~\cite{mi2021information,zhou2021operator}.
The models are obtained by mapping the unitary operator dynamics to a classical stochastic process by disorder average.
While this approach is feasible for usual noisy/random models without symmetry, adding charge conservation makes the problem more difficult as it causes quantum coherence to persist at the operator level, even after disorder-averaging (Fig.~\ref{fig:intro_fig}(e)). However, such a classical description is important to obtain for charged models, as thermalization in systems with charge conservation has recently become accessible on quantum simulators~\cite{Zhou_charged_2022}.

There is extensive literature on operator-dynamics~\cite{Patel_2017,Khemani_2018,Rakovszky_2018, hunter2018operator, chen2020quantum,Chen_2020,bao2021symmetry, agarwal2022emergent, gu2017energy, pai2019localization, feldmeier2021critically, cotler2020spectral, zanoci2022near} and spectral form factors~\cite{friedman2019spectral,Roy2020SFF, kos2021chaos,Moudgalya_2021,singh2021subdiffusion,roy2022spectral} in the presence of symmetries. The primary mechanism through which charge, or other conserved quantities, can influence operator growth is by confining the access of growing operators to a specific (symmetry) sector of the Hilbert space. 
In addition, within extended systems, the conservation law leads to transport of local conserved quantities, which couples to the operator dynamics. Therefore the correct semi-classical picture of operator dynamics in the presence of a symmetry should at least contain two dynamical variables, the operator size and the local density of the conserved quantity. 


In this work, we study operator dynamics in the presence of charge transport. 
We derive the required semi-classical equations which couple the charge and the operator, that are valid even in inhomogeneous and dynamical charge-density backgrounds:
\bea
\label{eq:FKPP_eqns}
&\partial_t \rho = \partial_r^2 \rho \\ 
&\partial_t \xi = \partial_r^2 \xi + 2 g^2 \xi (\xi^2 - \rho(1-\rho))
\eea
Here $\rho(r,t)$ is the charge density, which obeys the diffusion equation, while $\xi(r,t)$ is the analog of operator-size in models with symmetries that measures local scrambling of the operator. The dynamics of $\xi$ is described by a diffusion-reaction equation that depends on the dynamical charge density. In particular, the local density bounds the range of $\xi$ from 0 to $\sqrt{\rho(1-\rho)}$. We note that this class of equations has been independently studied in the context of bacterial population growth in diffusive media, where it is denoted as the `Diffusive Fisher–Kolmogorov equation'~\cite{golding1998studies}.

To derive these equations, we use the complex Brownian SYK model on a lattice with U(1) symmetry. SYK models, which are all-to-all interacting models with random couplings~\cite{Sachdev1992}, are well-known for their connection to black-hole physics~\cite{Sachdev_2015, kitaev2015} and for their exactly solvable nature in the large-$N$ limit~\cite{Maldacena_2016_SYK, Polchinski2016, Gross_2017, Gu_2020}. Their Brownian limit, which considers time-dependent couplings, has also been studied in various contexts~\cite{saad2018semiclassical, Sunderhauf2019, Chen_2020, jian2021measurement, Jian_2021, stanford2022subleading, agarwal2022emergent}. The dynamics of Brownian models can be mapped to a stochastic process~\cite{Xu_2019, Zhou_2019, Jian_2021}, or the imaginary time dynamics of bosonic-models~\cite{Sunderhauf2019}. It has also been shown that Brownian SYK models, in particular, give rise to emergent symmetry structures after the disorder averaging procedure which can be used to compute the OTOC both at large finite $N$ and in the infinite-$N$ limit~\cite{agarwal2022emergent}. 
In this work, we show that the extended complex Brownian model is mapped to a quantum SU(4) spin chain with inter-site Heisenberg coupling and intra-site interaction. In the large $N$ limit, the microscopic quantum model is reduced to the semi-classical equations in Eq.~\eqref{eq:FKPP_eqns}. We provide a complete picture of the coupled dynamics between operators and charge. Our approach can also be extended to systems with other symmetries.

The rest of the paper is organized as follows:
Sec.~\ref{sec:Summary} provides a brief summary of the main results.
In Sec.~\ref{sec:operator_basis}, we discuss the operator-basis which respects U(1) symmetry and can therefore be used to precisely characterize the four dynamical charges involved in the computation of the OTOC. In Sec.~\ref{sec:gen_pro}, we describe the details of the complex SYK model, and the mapping of the effective imaginary-time evolution in the Brownian model to an SU(4) spin-chain. In Sec.~\ref{sec:OperatorStateMap} we describe the mapping of operator-states of complex fermions to GT-patterns in the SU(4) algebra. We also map the physical charges to the weights of the algebra.
Sec.~\ref{sec:coherent_transitions} covers the primary distinguishing feature of operator dynamics restricted by a U(1) symmetry, namely the presence of transitions which introduce quantum operator coherence. We describe the procedure to compute the infinite-$N$ limit in Sec.~\ref{sec:large_N}, within which we also describe the general structure of the `restricted' Fokker-Planck equation obtained for quantum many-body models with charge conservation. In Sec.~\ref{sec:OTOC_large_N}, we derive the equation governing the OTOC, namely the charge-dependent FKPP equation. 
This reveals some interesting features of operator-dynamics in the presence of symmetries, as we show in section Sec.~\ref{sec:Domain_wall}, where we examine operator-dynamics in different charge domain-wall backgrounds and observe the strong influence of charge dynamics on even non-conserved operators.
Sec.~\ref{sec:conclusions} concludes this work with a brief summary and some discussions on a contrasting view of charge vs.\ energy conservation.


\section{Summary of the main results}
\label{sec:Summary}
\begin{figure}
    \centering    
    \includegraphics[width = \columnwidth]{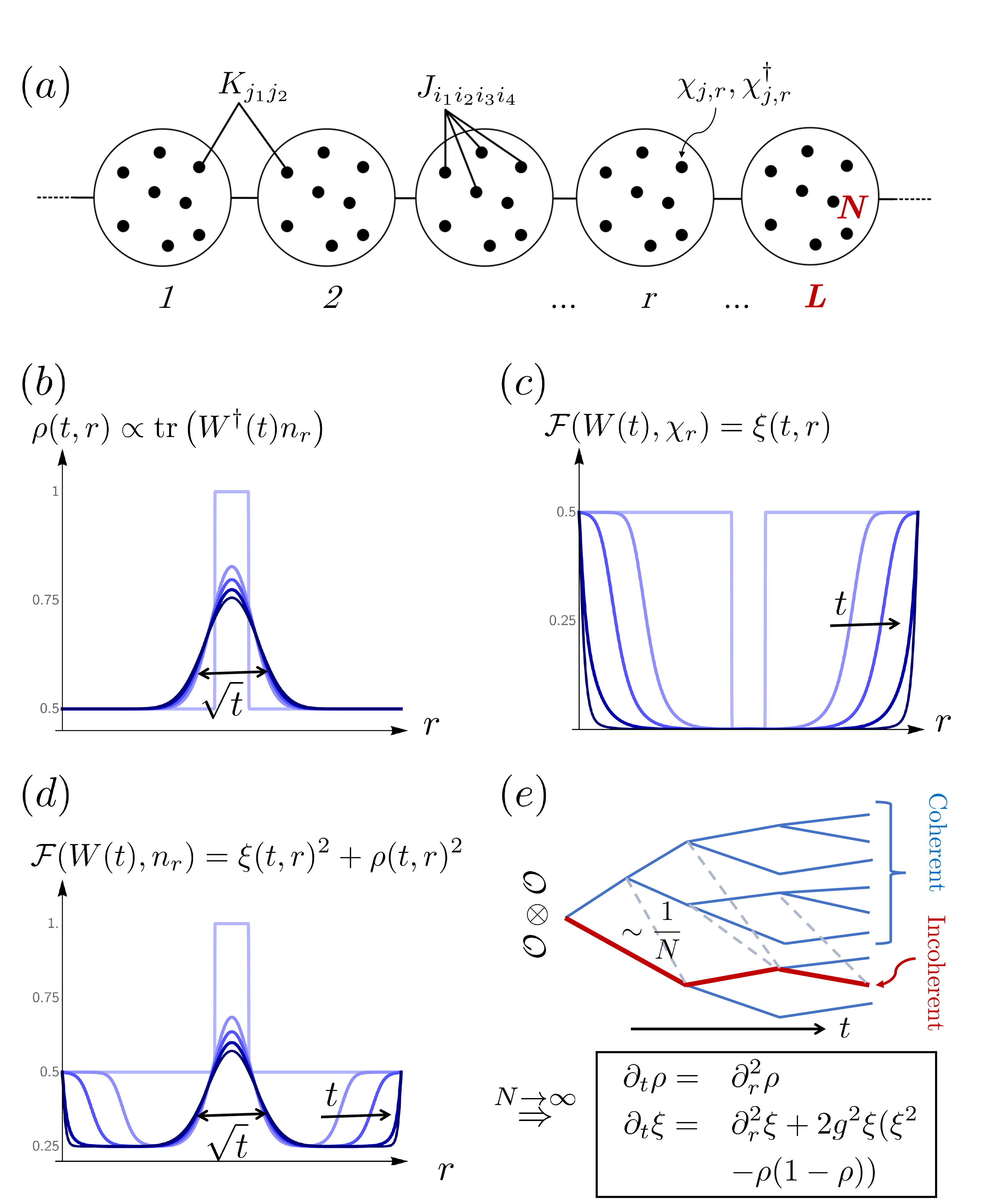}
    \caption{(a) The Brownian SYK model with $L$ clusters of $N$ complex fermions, with on-site interaction $J$ and nearest neighbor hopping $K$. The total fermion number operator $\sum n_{r,i}$ is the conserved operator, leading to a $U(1)$ symmetric circuit. (b-d) The time evolution of charge $\rho$ (b),the OTOC $\mathcal{F}(W(t),\chi_{r})$ (c), and OTOC $\mathcal{F}(W(t),n_{r})$ (d), when the initial operator $W$ has an initial charge distribution as shown in (b). In each of the graphs the darkness of the plots increases with increasing times. The charge (b) and conserved part of the operator (d) have diffusive behavior, while the OTOC $\mathcal{F}(W(t),\chi_{r})$ (c) and the uncharged part of OTOC $\mathcal{F}(W(t),n_{r})$ propagates ballistically. (e) Schematic of the time evolution of two copies of the operator, which is involved in the OTOC computation. Due to the $U(1)$ symmetry, the super-operator develops coherence, which makes the hydrodynamic description hard. In this work, we argue that for certain probing operators, the correction from this induced coherence is $1/N$ suppressed. In the $N\to\infty$ limit, we directly obtain the hydrodynamical equations~\ref{eq:FKPP_eqns}, which lead to the dynamical evolution in (b-d).} 
    \label{fig:intro_fig}
\end{figure}
The primary result of our work is captured by Eq~\eqref{eq:FKPP_eqns}. The equation  effectively models the evolution of the OTOC depicted in Eq.~\eqref{Eq:OTOC_definition}, for a charge-conserving fermionic model defined on a chain (Fig.~\ref{fig:intro_fig}(a)). The steps via which this connection is established are:
\begin{itemize}
\item To begin, one specifies the charge density on each site $r$. This fixes the profile of $\rho(r,t=0)$.
\item Because the initial charge on every site has been fixed, the simplest initial operator is the projection operator onto a symmetry sector with a given charge density on every site.
\item In models where the charge is conserved on each site, this projector is static. However, once the charge is allowed to flow from site to site, this operator becomes dynamical as well. 
\item Just as there are multiple states within each charge sector, there are also multiple operators~\cite{agarwal2022emergent}. The variable $\xi$ controls the choice of the operator once the charge is fixed. For example, the maximal value of $\xi=\sqrt{\rho(1-\rho)}$ represents a simple operator such as a projector, while $\xi=0$ represents a local scrambled operator within the charge sector. This is consistent with $\xi=0$ being a stable solution of Eq.~\eqref{eq:FKPP_eqns}, as all simple operators will eventually evolve into the most complex one.
\item Fixing both $\rho(t=0)$ and $\xi(t=0)$ also completely fixes the initial global operator $W(t=0)$ in Eq.~\eqref{Eq:OTOC_definition}. Hence the operator $W$ carries information about two separate modes, the charge, and the `complexity' of the operator within the charge subspace. Following this, the operator is evolved using Eq.~\eqref{eq:FKPP_eqns}, where it is observed that the mode $\xi$ spreads ballistically while $\rho$ is governed by diffusion.
\item The evolved operator is then measured via the use of a probing operator $V$. The choice of whether $V$ is chosen to be a conserved (has overlap with charge) or non-conserved operator determines which modes the OTOC detects:
\bea
&\F(t)|_{V=\chi_{r_0}} = \xi(r_0,t) \\ &\F(t)|_{V=n_{r_0}} = \xi^2(r_0,t) + \rho^2(r_0,t)
\eea
\end{itemize}

The operator $\chi_{r_0}$ refers to the local creation operator on site $r_0$ of the chain, while $n_{r_0}$ is the local number operator, which has overlap with the charge. This formalism distinguishes the time-ordered Green's function from the OTOC, as the former can only detect the charged mode and not the fixed-charge operator transitions (Fig.~\ref{fig:intro_fig}). It is important to mention that $W$ is a non-local operator as it fixes the initial charge profile (Sec.~\ref{sec:operator_basis}). This is necessary if one wishes to obtain a precise description of how information dynamics are related to charge dynamics, as local operators do not have a well-defined charge. This is perhaps also related to other works which have observed that locality has non-trivial implications in the presence of symmetries~\cite{marvian2022restrictions}. The equations in Eq.~(\ref{eq:FKPP_eqns}) correctly reproduce all known features of charged chaotic models, such as the charge-dependent Lyapunov exponent/butterfly velocity~\cite{Chen_2020,chen2020quantum}, and the late-time diffusive tail of the OTOC~\cite{Khemani_2018,Rakovszky_2018, cheng2021scrambling}. Moreover, since they are valid for arbitrary initial charge/operator profile, we simulate the coupled equations in inhomogeneous backgrounds to obtain new phenomena.

The formalism developed in this work encodes the microscopic operator dynamics in terms of transitions between states in a particular irrep of the SU(4) algebra. Since the OTOC is computed on four time-contours, this allows us to track the dynamics of the four corresponding conserved charges in terms of the weights of the SU(4) algebra, and explains the emergence of the coupled equations describing the OTOC in terms of a single charged mode ($\rho$) and a non-conserved ballistic mode ($\xi$). This derivation reveals new features of operator dynamics in the presence of conservation laws. Namely, in contrast with models that do not have continuous symmetries, the operator dynamics in the model with a U(1) symmetry allow for transitions that introduce quantum coherence at the operator level. Therefore, while usual noisy/random models are well described by a classical stochastic process, for U(1) symmetric models the time-evolution of the operator can only be modeled by a probability in a fixed subspace, and in the infinite-$N$ limit, this gives rise to a Fokker-Planck equation which only describes a conserved quantity in the `incoherent' sector of the operator-states.

\section{Charged operator basis}
\label{sec:operator_basis}

In this work we will be concerned with computing the OTOC, which can be viewed through the lens of operator spreading. Due to the charge conservation, we work with a specific choice of operator basis. We assume the degrees of freedom are represented by complex fermions, which satisfy the anti-commutation relations:
\[ \{ \chi_i^{\dagger}, \chi_j \} = \delta_{ij}   \; ; \; \{ \chi_i, \chi_j \} = 0 \] 
The relevant operators we consider are left and right eigen-operators with respect to the U(1) symmetry:
\bea \label{eq:charge_profile}
\bigg(\sum_i n_i \bigg) O = q_a O ; \quad O \bigg(\sum_i n_i \bigg) = q_b O
\eea
where $(n = \chi^{\dagger} \chi)$ is the number operator. It can be easily checked that for a model with U(1) conservation, the charges $(q_a,q_b)$ are a constant of motion during the dynamics of the operator string. For a single fermionic operator, we choose the relevant four dimensional eigen-operator basis and the charges take the following respective values:
\bea \label{eq:singleoperatorcharge}
\chi^\dagger:(1, 0) \ \ \chi:(0,1) \ \ n:(1,1) \ \ \bar{n}:(0, 0)
\eea
where $\bar{n} = I-n$. Following this insight, we will work with operator strings $\mathcal{S}$ when the system contains $N$ fermions, where each element of the string is picked from the U(1) operator basis defined above: 
\bea
\mathcal{S} = 2^{N/2}s_1 s_2 \cdots s_N, \quad s_i \in \{ \chi_i^\dagger, \chi_i, n_i , \bar{n}_i \}.
\eea
This ensures that the entire operator string is also an eigen-operator of the global U(1) symmetry. The factor of $2^{N/2}$ in the string is picked to ensure the following orthogonal and completeness relations
\bea
\frac{1}{\tr I} \tr (\S^\dagger \S') =  \delta(\S',\S), \ \ \frac{1}{\tr I} \sum\limits_{\S} \S^{\dagger}_{mn} \S_{pq} = \delta_{mq}\delta_{np}.
\label{eq:orthcomp}
\eea

\begin{figure}
\includegraphics[width=1\columnwidth]{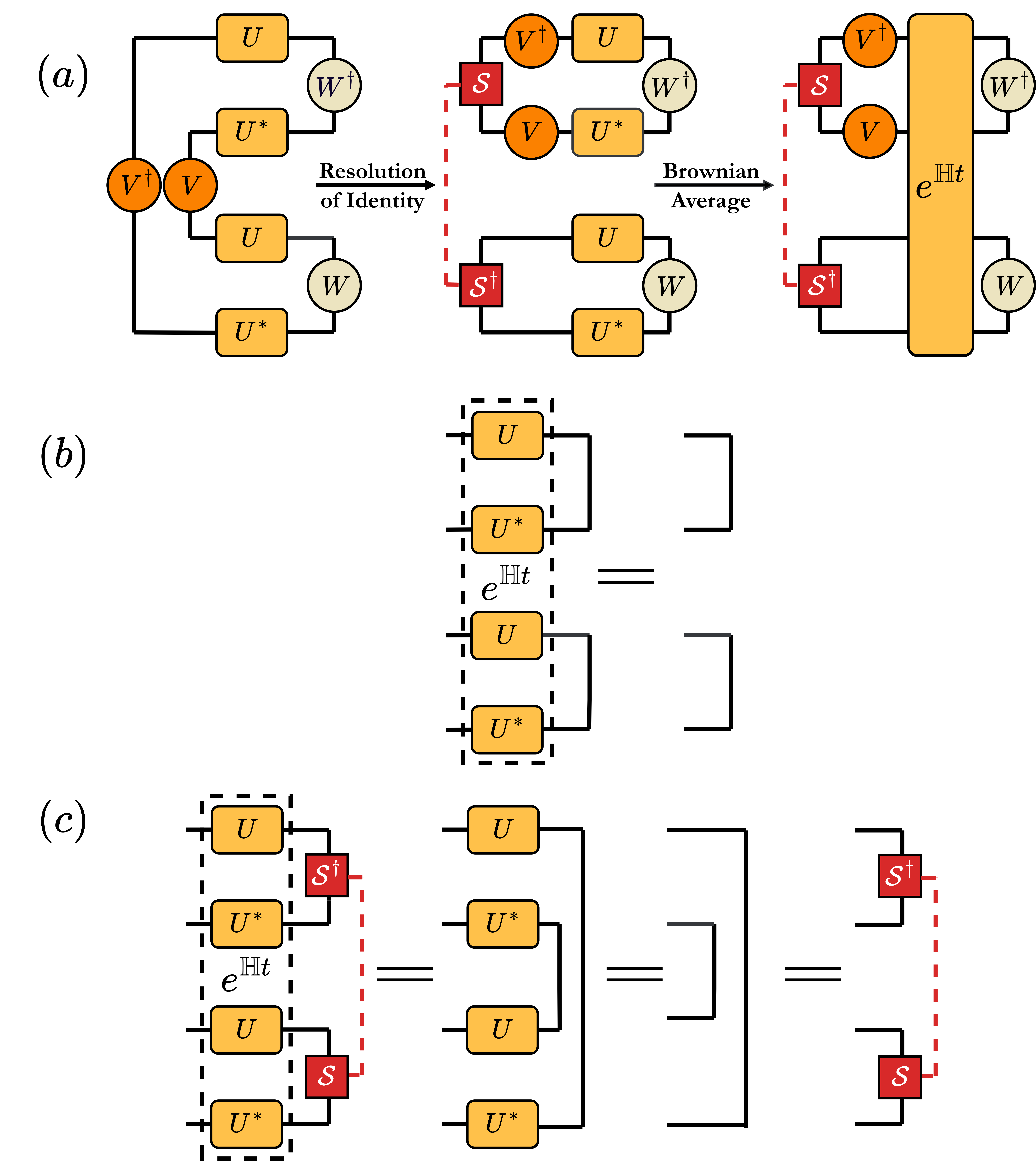}
\caption{Figure adapted from~\cite{agarwal2022emergent}. In the first part of (a), the process of resolving the identity in the output state to ensure a proper overlap with the input state is depicted. For the Brownian model, disorder-averaging after this procedure turns the OTOC into an amplitude, computed after the states are evolved in imaginary time using an effective Hamiltonian $\H$. (b) and (c) depict the null-eigenstates on four-time contours, which are a consequence of the dynamics being Unitary. While (b) is simply the double-copy of the identity, (c) represents the fully-scrambled steady state $\sum_{\S} \ket{\S^{\dagger} \otimes \S}$ if the dynamics are also chaotic.}
\label{Fig:OTOC-Tensor}
\end{figure}

The procedure to compute the OTOC, on the other hand, begins by rearranging the correlator to write it in the operator state language:
\bea \label{eq:OTOC}
\F(W(t),V) &= \frac{1}{\tr{I}} \tr{(W^{\dagger}(t) V^{\dagger} W(t) V)} =  \tr{I}\bra{\o} \mathbb{U} \ket{\i}\\
\ket{\i} &=\frac{1}{\tr I}\sum W^\dagger_{m n} W_{pq}\ket{m\otimes n \otimes p \otimes q} \\
\ket{\o} &= \frac{1}{\tr I}\sum V^\dagger_{m q} V_{p n }\ket{m\otimes n \otimes p \otimes q}\\
\mathbb U &= U\otimes U^* \otimes U \otimes U^*
\eea
Hence the computation of the OTOC involves four copies of the Unitary and in a charge-conserving model, four corresponding conserved charges as well. These conserved charges, which we label as $(q_a, q_b, q_c, q_d)$, are the left and right charges of the double-copy of the operator-state involved in the OTOC. Therefore, an exact dynamical description of the OTOC would depend on the evolution of these four independent charges as well.

To complete the formalism and ensure a proper overlap between the input and output state, we insert the resolution of identity in the output state (Fig.~\ref{Fig:OTOC-Tensor}(a))
\bea
\ket{\o} &=\frac{1}{\operatorname{tr}^{2} I} \sum V_{m m^{\prime}}^{\dagger} \mathcal{S}_{m^{\prime} n^{\prime}}^{\dagger} V_{n^{\prime} n} \mathcal{S}_{p q}|m \otimes n \otimes p \otimes q\rangle \\
&=\frac{1}{\operatorname{tr}^{2} I} \sum_{\mathcal{S}}\left|V^{\dagger} \mathcal{S}^{\dagger} V \otimes \mathcal{S}\right\rangle
\eea

\section{General procedure and emergent SU(4) algebra}
\label{sec:gen_pro}

The model we work with is the complex Brownian SYK chain, defined through the all-to-all quartic on-site and quadratic two-site interaction Hamiltonians
\bea \label{eq:Hamiltonians}
H_{\text{intra}} &= g \sum_{i_1,i_2,i_3,i_4,r} J^{r}_{i_1,i_2,i_3,i_4}(t) \chi^{\dagger}_{i_1,r} \chi^{\dagger}_{i_2,r} \chi_{i_{3},r}\chi_{i_{4},r} + \text{h.c.} \\ 
H_{\text{inter}} &= \sum_{j_1,j_2,r} K^{r}_{j_1,j_2}(t) \chi^{\dagger}_{j_1,r} \chi_{j_2,r+1} + \text{h.c}.
\eea 
where $i,j,k,l$ are indices labelling fermions on each site and $r$ is the site/cluster index. Each cluster contains $N$ fermions and the number of clusters can be chosen to be $L$. Although we consider the one-dimensional model, the formalism is easily generalized to higher dimensions. The couplings in the Brownian model break the time-translation symmetry and satisfy the constraints
\bea
\overline{J^{r}_{i_1,i_2,i_3,i_{4}}(t) J^{r'*}_{j_1,j_2,j_3,j_4}(t')} &=\frac{1}{N^3} \delta_{i_1,j_1}...\delta_{i_{4},j_{4}} \delta_{r,r'} \delta(t-t')\\
\overline{K^{r}_{i_1,i_2}(t) K^{r'*}_{j_1,j_2}(t')} &= \frac{1}{N}\delta_{i_1,j_1} \delta_{i_{2},j_{2}} \delta_{r,r'} \delta(t-t')
\eea
Hence the couplings are uncorrelated at different times, and this property makes the Brownian model more analytically and numerically tractable. The on-site Hamiltonian discussed above conserves charge on each site, whereas the inter-site term facilitates the flow of charge between clusters but preserves the global charge. The inter-site model is also chosen to be non-interacting, in contrast to its intra-site counterpart which scrambles quantum information within each cluster. In this work we will precisely explore this interplay between transport and scrambling.

After disorder averaging, the composite unitary $\mathbb U$ in Eq.~(\ref{eq:OTOC}) for the Brownian model is mapped to an emergent Hamiltonian $\mathbb H$ which evolves in imaginary time
\bea 
\overline{\mathbb U}  = \overline{U^a  U^{b*} U^c  U^{d*}} = e^{\mathbb H t}.
\eea
Here we have labelled individual time-contours by the labels $a,b,c,d$. For complex fermionic models, after the disorder average, each copy is occupied by anti-commuting fermions after a Jordan-Wigner like transformation
\bea
\{ \psi^{\dagger \alpha}_{i,r},\psi_{j,r'}^{\beta} \} = \delta_{i,j} \delta_{\alpha, \beta} \delta_{r, r'}  ; \quad  \alpha, \beta=a,b,c,d
\eea
The details of this transformation are present in the appendix. An important point to note is that we have performed a particle-hole transformation on replicas $b$ and $d$, for future mathematical convenience. While for general fermionic models this procedure does not make computations simpler, one can show that for the Brownian SYK models considered in this work, the emergent Hamiltonian can be written purely in terms of permutation symmetric inter and intra replica number operators which are the generators of the SU(4)$\otimes$U(1) algebra:
\bea \label{eq:emergent_algebra}
&\H = \H(S_r^{\alpha \beta}) ; \quad S_r^{\alpha \beta} = \sum_{i}\psi^{\dagger \alpha}_{i,r} \psi_{i,r}^{\beta} , \quad  \alpha, \beta=a,b,c,d\\
&[ S_{r}^{\alpha \beta}, S_{r}^{\gamma \sigma} ] = \delta^{\beta \gamma} S_r^{\alpha \sigma} - \delta^{\alpha \sigma} S_r^{\beta \gamma}.
\eea
The 16 generators $S_r^{\alpha \beta}$ can be split into the 15 generators of SU(4) plus one given by $\sum_{\alpha} S_r^{\alpha \alpha} = Q_r$ which forms the U(1) part of the algebra and commutes with the SU(4) generators. Hence we have exactly mapped the dynamics of a Brownian all-to-all interacting model to the dynamics of an SU(4) spin chain that evolves in imaginary time. As an example, for the free-model, i.e. $H_{\text{inter}}$ term in Eq.~\eqref{eq:Hamiltonians}, the emergent Hamiltonian takes the form of the SU(4) Heisenberg model:
\bea
\H_{\text{inter}}= \frac{1}{N} \bigg( \sum_{\alpha,\beta,r} S^{\alpha \beta}_r S^{\beta \alpha}_{r+1}-N \sum_r Q_r \bigg)
\eea
This has an appealing format, as it makes the global SU(4) invariance of the emergent model manifest. Models that have interactions, on the other hand, will lack this exact symmetry. However, if the model has global charge conservation, it will be reflected in terms of global weight conservation in the emergent SU(4) spin model. In other words, while the emergent Hamiltonian corresponding to the free model commutes with all global SU(4) operators and is therefore SU(4) symmetric, more general models, which conserve charge but have interactions, will only commute with the Cartan-subalgebra and conserve weights.

Physically, the structure of charges in terms of weights can be understood as follows: Each replica has its own copy of the globally conserved charge labelled by $\sum_r(S_r^{aa},N-S_r^{bb},S_r^{cc},N-S_r^{dd})=\sum_r (q_{a,r},q_{b,r},q_{c,r},q_{d,r})$ respectively, where the $S_r^{\alpha \alpha}$ operators are defined in Eq.~\eqref{eq:emergent_algebra}. The physical charge is counted differently on replicas $b,d$ due to the particle-hole transformation (App.~\ref{App:four_time_fermions}). On the right hand side of the equation, $q_r$'s are essentially the left and right charges of the two copies of the operator on each site, as defined in Eq.~\eqref{eq:charge_profile}. However, not all four of them are independent as the Brownian nature of the model fixes the sum $\sum_{\alpha} S_r^{\alpha \alpha}= Q_r$, which forms the U(1) part of the algebra. For example, for the irrep $(0,N,0)$, we have a fixed $Q=2N$ as it commutes with all generators of the SU(4) algebra. The remaining three charges can be mapped to the three weights of the SU(4) algebra, which will be made more precise in the next section.

\begin{figure}
\includegraphics[width=1\columnwidth]{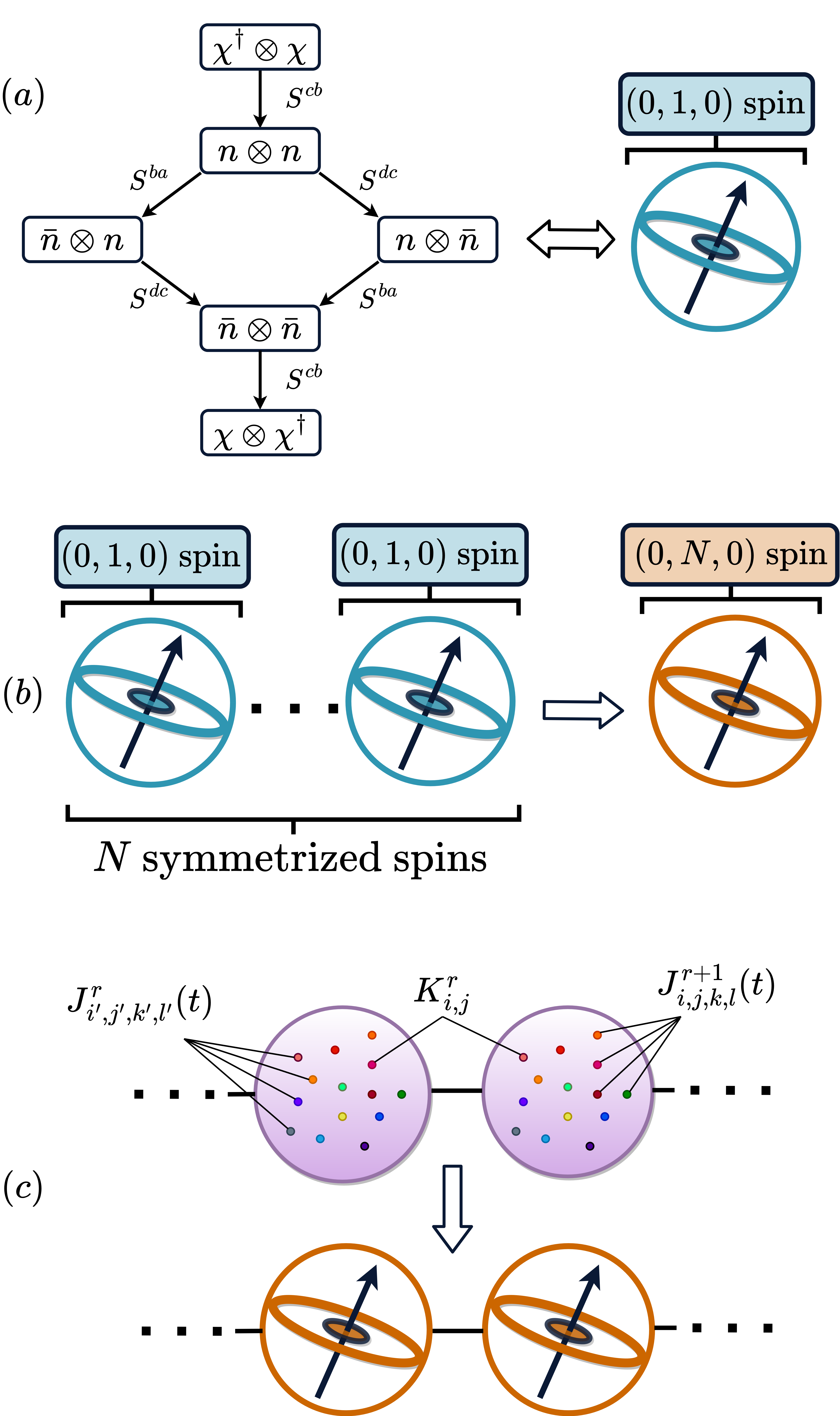}
\caption{(a) The SU(4) representation $(0,1,0)$ for the relevant operator-states corresponding to a single fermion, along with the rules that govern transitions between the operator-states. Here $\bar{n} = I - n$ is the operator orthogonal to the number operator $n$. (b) The emergence of the SU(4) spin $(0,N,0)$ as the permutation-symmetric combination of $N$ fermions within each cluster in the Brownian SYK chain. (c) After disorder average, each cluster in the chain maps to a $(0,N,0)$ spin, and in this work we consider the limit of infinite-$N$.}
\label{Fig:Complex-Spin}
\end{figure}

\section{Map from operator strings to SU(4) basis}
\label{sec:OperatorStateMap}
For a single complex fermion, the four copies $a,b,c,d$ together span a 16 dimensional Hilbert space, which can be thought of as the double-copy of four-dimensional operator basis. For the Brownian model, as discussed in the previous section, the emergent Hamiltonian is composed of SU(4) operators, which suggests that the corresponding states should be viewed in terms of the representations of the algebra. The 16 states split into irreps as:
\bea
\textbf{16} = \textbf{1} \oplus \textbf{4} \oplus \textbf{6} \oplus \bar{\textbf{4}} \oplus \textbf{1}.
\eea
In this work, we will be concerned with the $\textbf{6}$ or in fundamental weight notation, the $(0,1,0)$ irrep, as it contains the operators of interest. The states in this irrep are listed in Fig.~\ref{Fig:Complex-Spin}(a), along with how the SU(4) operators act on them. This six-dimensional operator space is similar to the one presented in~\cite{Rakovszky_2018}.

Now we will move on to the discussion of representations which involve multiple fermions. As is well known, for SU($n$) with $n>2$, there are $n-1$ weights and multiple states with the same weights in an irrep, hence states cannot be uniquely labeled by just using the Casimir and the weights. To study the large-$N$ dynamics, we will therefore begin by discussing the basis labeling the states on each site in the SU(4) spin chain. The scheme to label the states relies on Gelfand-Tsetlin patterns, where 10 numbers are used to label each state within each SU(4) irrep~\cite{Alex_2011}: 
\[ \begin{pmatrix} m_{1,4}&&m_{2,4}&&m_{3,4}&&m_{4,4}\\&m_{1,3}&&m_{2,3}&&m_{3,3}&\\&&m_{1,2}&&m_{2,2}&&\\&&&m_{1,1}&&&\end{pmatrix}\]
The numbers have to satisfy the constraint: \[m_{k, l} \geq m_{k, l-1} \geq m_{k+1, l}\] in order to represent a valid GT-pattern. In this work, we will fix the irrep on each cluster to be $(0,N,0)$, which is the unique permutation symmetric tensor composition of $N$ $(0,1,0)$ irreps (Fig.~\ref{Fig:Complex-Spin}(b)).
\bea
    \underbrace{(0,1,0) \otimes...\otimes (0,1,0)}_{N \text{ times}} = (0,N,0) \oplus \underbrace{(1,N-2,1)}_{N-1 \text{ copies}} \oplus ...,
\eea
Physically, this corresponds to restricting the operator-strings of $N$ fermions on each cluster to be permutation symmetric among the indices which label the fermions. This is justifiable for two reasons, firstly, it is expected that in the case of scrambling dynamics, the non-symmetric operator-strings will decay exponentially fast, leaving behind only the symmetric strings after a short time. Secondly, it has also been shown that features such as Lyapunov growth and the scrambled steady-state are entirely contained within this symmetric irrep, making it the only relevant sector to study the chaotic behavior of the model~\cite{agarwal2022emergent}. For the Brownian model, starting from a symmetric operator-string completely restricts the dynamics to the symmetric sector, which can be seen from the emergent algebra structure.

Fixing the irrep also completely fixes the first row of the GT-pattern to $(m_{1,4}, m_{2,4}, m_{3,4}, m_{4,4}) = (N,N,0,0)$. Along with this, the entries $m_{1,3} = N, m_{3,3} = 0$ are also fixed due to the constraints of the GT-pattern. This leaves us with four variables on each site $r$, which can be understood in terms of the SU(4) algebra as the three independent weights $w_{1,r},w_{2,r},w_{3,r}$ plus an index $n_r$ which labels the state within each fixed-weight subspace.
\begin{equation} \label{Eq:GT-Patttern}
    \begin{pmatrix} 
    N&&N&&0&&0\\&N&&w_{1,r}&&0&\\&&w_{2,r}-n_r&&n_r&&\\&&&w_{3,r}&&& \end{pmatrix}     \equiv \ket{w_{1,r},w_{2,r},w_{3,r},n_r}
\end{equation}
As has been discussed in the previous section, the four emergent charges on the four time-contours $(q_{a,r},q_{b,r},q_{c,r},q_{d,r})$ map to the three weights of the SU(4) algebra plus one overall total charge, which is not dynamical once the irrep is fixed. The variables $w_{1,r}, w_{2,r}, w_{3,r}$ encode the three remaining dynamical charges, and the variable $n_r$ controls the fluctuations within each fixed-weight sector, i.e.\ it represents the dynamics of the operator within a fixed charge subspace. The weights are related to the physical charges on the four contours in the following way:
\bea \label{eq:charges_weights}
&q_{a,r} = w_{3,r} \, ; \, q_{b,r} = N-w_{2,r}+w_{3,r} \\
&q_{c,r} = N-w_{2,r}+w_{1,r} \, ; \, q_{d,r} = w_{1,r}
\eea

\section{Quantum operator coherence due to U(1) symmetry}
\label{sec:coherent_transitions}
This section will be devoted to the comparison of operator dynamics in Brownian models with and without charge conservation. An example of a model without charge conservation is the Brownian SYK model built with Majorana fermions. The operator transitions in such a model strictly follow the general rule:
\bea
\ket{O_1^{\dagger} \otimes O_1} \rightarrow \ket{O_2^{\dagger} \otimes O_2} 
\eea
Such transitions can be seen as `incoherent' when translating from the operator state to super-operator language:
\bea
\ket{O_1 \otimes O^{\dagger}_2} \rightarrow \ket{O_1} \bra{O_2}
\eea
which means that diagonal elements in the super-operator can only transition to other diagonal elements. This general rule is also compatible with the parity symmetry in the Majorana model, which restricts all operators of the form $\ket{O^{\dagger} \otimes O}$ to one symmetry sector since all such operators have even parity~\cite{agarwal2022emergent}.

On the other hand, in the complex model, only a part of the symmetry-resolved emergent Hilbert space consists of incoherent operator-states (Fig.~\ref{Fig:Complex-Spin}). Specifically, the operator-states $\ket{\bn \otimes n}$ and $\ket{n \otimes \bn}$ break the incoherence. Requiring this incoherence in the operator-state forces the charges to obey the necessary constraints $q_{a,r} = q_{d,r}, q_{b,r} = q_{c,r} \implies w_{1,r}=w_{3,r}$, which in general is not true for a random state in the Hilbert space. One can use the information in Fig.~\ref{Fig:Complex-Spin} to reproduce the following transition generated by a global-charge conserving operator $S_1^{ca}S_2^{ac}$ on an operator string over 2 fermions:
\bea
S_1^{ca}S_2^{ac} \ket{\chi_1^{\dagger} \chi_2 \otimes \chi_1 \chi_2^{\dagger} } \rightarrow  \ket{ \bn_1 n_2 \otimes n_1 \bn_2}.
\eea
This can be viewed as a toy model with two sites and one fermion per site. The operator $S_1^{ca}S_2^{ac}$ is one of the terms in the emergent Hamiltonian corresponding to the hopping term in Eq.~\eqref{eq:Hamiltonians}, which conserves charge over the two fermions. Even though the transition conserves the four global charges $(q_a, q_b, q_c, q_d) = (1,1,1,1)$ of the entire operator-string, it introduces operator coherence on each fermion, which means the operators on the right hand side cannot be written in the form $\ket{O^{\dagger} \otimes O}$ and the corresponding super-operator develops off-diagonal entries. This insight distinguishes the cases with and without charge conservation, and also explicitly affects the large-$N$ hydrodynamics of the operators. In this work, we will utilize permutation symmetric operator-states on each site, which are chosen from the irrep $(0,N,0)$, and labelling the states as incoherent is equivalent to the condition $w_{1,r} = w_{3,r}$.

\section{The large-$N$ formalism and the `restricted' Fokker-Planck equation}
\label{sec:large_N}
The strategy to obtain the large-$N$ hydrodynamical equations of motion relies on a similarity transformation that turns the effective Hamiltonian $\H$ into a (partially) stochastic matrix. As can be seen from Fig.~\ref{Fig:OTOC-Tensor}, the operator-state $\sum_{\S}\ket{\S^{\dagger} \otimes \S}$ is an exact eigenstate of $\H$ due to the unitary nature of the underlying dynamics:
\bea
\H \sum_{\S}\ket{\S^{\dagger} \otimes \S} = 0.
\eea
This steady-state represents the late-time distribution of an initially simple operator under scrambling dynamics. For non-conserved local operators which have overlap with this state, all other states which are orthogonal to the steady-state decay to zero at late time, due to the chaotic nature of the Brownian model. The steady-state can be resolved into different charge sectors, where each component acts as a scrambled steady-state for the specific charge sector (App.~\ref{App:Identity}). Thus, when starting from a specific charge sector, the operator evolves into the charge-resolved steady-state at late times, which also signals the decay of the OTOC. The full steady state on the chain can be written as a product state of the steady state on every site $r$:
\bea
\sum_{\S}\ket{\S^{\dagger} \otimes \S} = \bigotimes \sum_{\S_r}\ket{\S_r^{\dagger} \otimes \S_r}
\eea
In the GT-pattern basis, this steady-state on each site is written as:
\bea \label{eq:steady_state}
&\sum_{\S_r}\ket{\S_r^{\dagger} \otimes \S_r} = \sum_{w_{1r},w_{2r},n_r} c_{w_{2r},n_r} \ket{w_{1r},w_{2r},w_{3r}=w_{1r},n_r} \\
&c_{w_{2r},n_r} = \sqrt{\frac{(w_{2r}+1-2n_r)}{N+1}\binom{N+1}{N-(w_{2r}-n_r)} \binom{N+1}{n_r}}
\eea
One can understand the restriction $w_{1r}=w_{3r}$ here from the perspective that the steady state $\sum_{\S_r}\ket{\S_r^{\dagger} \otimes \S_r}$ is strictly within the incoherent subspace of the larger Hilbert space. However, fixing $w_{1r}=w_{3r}$ is merely the necessary condition to obtain the state, whereas the sufficient condition involves also specifying the correct coffecients $c_{w_{2r},n_r}$. Following this, we define a diagonal matrix $\mathbb{S}_r$ with the entries:
\bea
\bra{w_{1r},w_{2r},w_{3r},n_r}\mathbb{S}_r\ket{w_{1r},w_{2r},w_{3r},n_r} = c_{w_{2r},n_r}
\eea
We use the matrix $\mathbb{S}_r$ to perform a similarity transformation in Eq.~\eqref{eq:OTOC}
\bea
\bra{\o} e^{\H t} \ket{\i} &= \bra{\tilde{\o}} e^{\tilde{\H} t} \ket{\tilde{\i}}\\
\tilde{\H} = \mathbb{S} \H \mathbb{S}^{-1} \, ; \, \ket{\tilde{\i}} &= \mathbb{S} \ket{\i} \, ; \, \ket{\tilde{\o}} = \mathbb{S}^{-1} \ket{\o}
\eea
The global transformation $\mathbb{S}$ is the tensor product of local transformations $\mathbb{S}_r$. This procedure has a two-fold advantage. Firstly, it removes the non-uniform $N$-dependence usually contained in the output state. As an example, if the probing operator $V$ in the OTOC is chosen to be the identity, the output state is the resolution of identity, meaning the steady state in Eq.~\eqref{eq:steady_state}. Hence the output state will contain combinatorial factors, which are not compatible with a continuum analysis. The Similarity transformation strips away these factors and makes the output state amenable to a large-$N$ expansion. Secondly, due of the choice of the Similarity matrix $\mathbb{S}$, the coefficients of the steady state corresponding to the resolution of identity (from the left) all become $1$, and hence the transformed Hamiltonian obeys the equation
\bea
\sum_{\vec{w}_1,\vec{w}_2,\vec{n}} \tilde{\H}_{(\vec{w}_1,\vec{w}_2,\vec{w}_3=\vec{w}_1,\vec{n}), (\vec{w}_1',\vec{w}_2',\vec{w}_3'=\vec{w}_1',\vec{n}')} = 0 
\eea
due to having a uniform left eigenvector. Here the vectors $\vec{w}_i, \vec{n}$ have components $(w_{i,r},n_r)$ over the space of all sites $r$. Therefore the matrix $\tilde{\H}$ has become stochastic, but only in the sector $\vec{w}_1=\vec{w}_3$ that contains the full steady state. This can be contrasted with the Majorana and spin version of the model, where the steady state occupies the entire Hilbert space and therefore the entire emergent Hamiltonian can be made stochastic~\cite{Xu_2019,chen2019quantum,Zhou_2019,agarwal2022emergent}. The stark difference between these two scenarios is precisely a consequence of the U(1) symmetry in the complex model.
We write the inner product which governs the OTOC in the wavefunction notation:
\bea
&\bra{\tilde{\o}} e^{\tilde{\H} t} \ket{\tilde{\i}} = \\&\sum_{\vec{w}_1,\vec{w}_2,\vec{w}_3,\vec{n}} \psi_{\o}(\vec{w}_1,\vec{w}_2,\vec{w}_3,\vec{n}) \psi_{\i}(\vec{w}_1,\vec{w}_2,\vec{w}_3,\vec{n},t)
\eea
where the input state is evolved using the emergent Hamiltonian
\bea \label{eq:input_evolution}
&\partial_t \psi_{\i}(\vec{w}_1,\vec{w}_2,\vec{w}_3,\vec{n},t) =\\ &\sum_{\vec{w}_1',\vec{w}_2',\vec{w}_3',\vec{n}'} \tilde{\H}_{(\vec{w}_1,\vec{w}_2,\vec{w}_3,\vec{n}),(\vec{w}_1',\vec{w}_2',\vec{w}_3',\vec{n}')} \psi_{\i}(\vec{w}_1',\vec{w}_2',\vec{w}_3',\vec{n}',t)
\eea
Due to the (partially) stochastic nature of $\tilde{\H}$, $\psi_{\i}$ behaves as a probability distribution, but only in the incoherent sector, and therefore
\bea \label{eq:restricted_probability}
\partial_t \big(\sum_{\vec{w}_1,\vec{w}_2,\vec{n}} \psi_{\i}(\w_1,\w_2,\w_3=\w_1,\n,t) \big) = 0
\eea
This will be explicitly reflected in the large-$N$ equation we obtain. We now follow the standard operating procedure and expand the SU(4) operators in $\H$, in the large-$N$ limit. The explicit matrix elements of simple raising/lowering operators in SU(n) are known in the GT-pattern basis~\cite{Alex_2011}.
For the purpose of expanding the operators, we use the variables
\bea \label{eq:continuum_labels}
&w_{1r}-w_{3r} \rightarrow 2 N v_r, \quad w_{2r} \rightarrow N y_r, \\ &w_{1r}+w_{3r}  \rightarrow 2 N \rho_{r}, \quad n_r \rightarrow N u_r.
\eea
The resulting equation, which is the continuum version of Eq.~\eqref{eq:input_evolution}, can be written in the form of a  `restricted' Fokker-Planck as follows:
\bea 
\partial_t \psi_{\i} = &\sum_{r}   \beta_r \psi_{\i} + \alpha_{v_r} \partial_{v_r}\psi_{\i} +  \partial_{u_r}(\alpha_{u_r} \psi_{\i})\\
& +\partial_{\rho_r}(\alpha_{\rho_r} \psi_{\i}) + \partial_{y_r}(\alpha_{y_r} \psi_{\i})
\eea
The explicit expressions for the functions $\alpha, \beta$ depend on the underlying model, but the `restricted' nature of the equation is always valid. As one can immediately check, $\int \psi_{\i}$ is not conserved due to the $\beta_r$ and $\alpha_{v_r}$ terms as they are not total derivative terms \footnote{For models that conserve charge on each site, there is no $\alpha_{v_r}$ term due to the corresponding weight conservation.}. However, they satisfy the important property:
\bea
\beta_r|_{\vec{v}=0} = 0, \quad \alpha_{v_r}|_{\vec{v}=0} = 0
\eea
for all $r$. Therefore, $\int \delta(\vec{v}) \psi_{\i}$ is conserved, which is expected from Eq.~\eqref{eq:restricted_probability} because $\vec{v}=0$ restricts to the incoherent sector. In what follows, we can restrict our analysis to this sector by choosing an output state strictly contained in it, since the overlap (and therefore the OTOC) will be blind to growth of the input state outside the sector. For example, choosing $V=\chi_{r_{0}} \footnote{We suppress the fermionic index and only display the cluster/site index since it is assumed that this operator will be restricted to the symmetric irrep, and therefore will be symmetrized between all the fermions in the cluster.}$  will satisfy this criterion, since $v_{r_0} = 1/N$ for the resulting output state and therefore it is in the sector to leading order in the infinite-$N$ limit. Hence we impose the condition $\vec{v}_r=0$ on the `restricted' Fokker-Planck by starting with an input state sufficiently well localized around $\vec{v}=0$ and obtain Langevin equations for the variables $y_r, \rho_r, u_r$ (Appendix ~\ref{App:Restr_FP}). Since the input state is chosen to be localized functions which evolve according to these Langevin equations, the OTOC will be given by:
\bea
\F(t) = \psi_{\o}(y_r(t),\rho_r(t),u_r(t))
\eea

\section{The OTOC at large $N$}
\label{sec:OTOC_large_N}
This section will be devoted to the study of the OTOC in the large-$N$ limit. We will probe various cases of the OTOC, such as when the dynamics are driven by interacting vs non-interacting Hamiltonians. We will also study how the behavior of the OTOC changes with the choice of operators involved. Before we discuss specific models, it is beneficial to recast the variables $y_r, \rho_r$ in terms of the more `physical' left and right charges $\rho_a$ and $\rho_b$ defined in Sec.~\ref{sec:operator_basis}:
\bea
\rho_{a,r} = \rho_r \; ; \; \rho_{b,r} = 1- y_r + \rho_r
\eea
This relation is true in the incoherent sector ($w_{1,r}=w_{3,r}$) as can be seen from the equations Eq.~\eqref{eq:charges_weights} and Eq.~\eqref{eq:continuum_labels}. Choosing the probing operator as $V= \chi_{r_0}$ means that the output state after the similarity transformation and in the continuum limit becomes:
\bea
\label{eq:cord_trans}
\sqrt{(1-\rho_{b,r}-u_{r})(\rho_{a,r}-u_{r})} = \xi_{r} \, ; \, \psi_{\o} = \xi_{r_0}
\eea
where we transform to the variable $u_r \rightarrow \xi_r$ at every site, in which case the OTOC is captured by the variable $\F(t) = \xi_{r_0}(t)$. The input operator $W$ on the other hand, is determined by fixing the initial value of $\xi_r, \rho_{a,r}$ and $\rho_{b,r}$ on every site.

\subsection{Free-fermionic chain}
In this section we will discuss the large-$N$ analysis for the free-fermion complex brownian SYK chain, namely the $H_{\text{inter}}$ term in Eq.~\eqref{eq:Hamiltonians}. For this strictly non-interacting case, the emergent Hamiltonian is simply the SU(4) Heisenberg model:
\bea
\H_{\text{inter}}= \frac{1}{N} \bigg( \sum_{\alpha,\beta,r} S^{\alpha \beta}_r S^{\beta \alpha}_{r+1}-N \sum_r Q_r \bigg)
\eea
and is therefore SU(4) invariant. This property can be used to exactly solve the OTOC even at finite $N$ (Appendix~\ref{App:free_fermion}). 
In the new co-ordinates $\xi, \rho_a, \rho_b$, we write down the restricted Fokker-Planck for the model and the corresponding equations of motion (Langevin equation) are:
\bea
\partial_t \rho_a =  \partial_r^2 \rho_a ; \quad \partial_t \rho_b =   \partial_r^2 \rho_b ; \quad \partial_t \xi =  \partial_r^2 \xi
\eea
As a reminder, the charges $\rho_a$ and $\rho_b$ are the left and right continuum charge densities of the charges $q_a$ and $q_b$ in Eq.~\eqref{eq:charge_profile} and $\xi$ measures the OTOC.
Here we have treated $r$ as a continuous variable. Notice that since we have restricted to the incoherent sector, the four emergent charges on the time-contours of the OTOC are not all independent and have collapsed into just two separate charges. We see some important features of the free fermionic chain from these equations of motion. The first observation is that the charges $\rho_a, \rho_b$ completely decouple from the OTOC $\xi$. Secondly, all three quantities show diffusive behavior, which is expected for the two charges from the global charge conservation, and for the OTOC due to the non-interacting nature of the underlying model. 

\begin{figure}
\includegraphics[width=1\columnwidth]{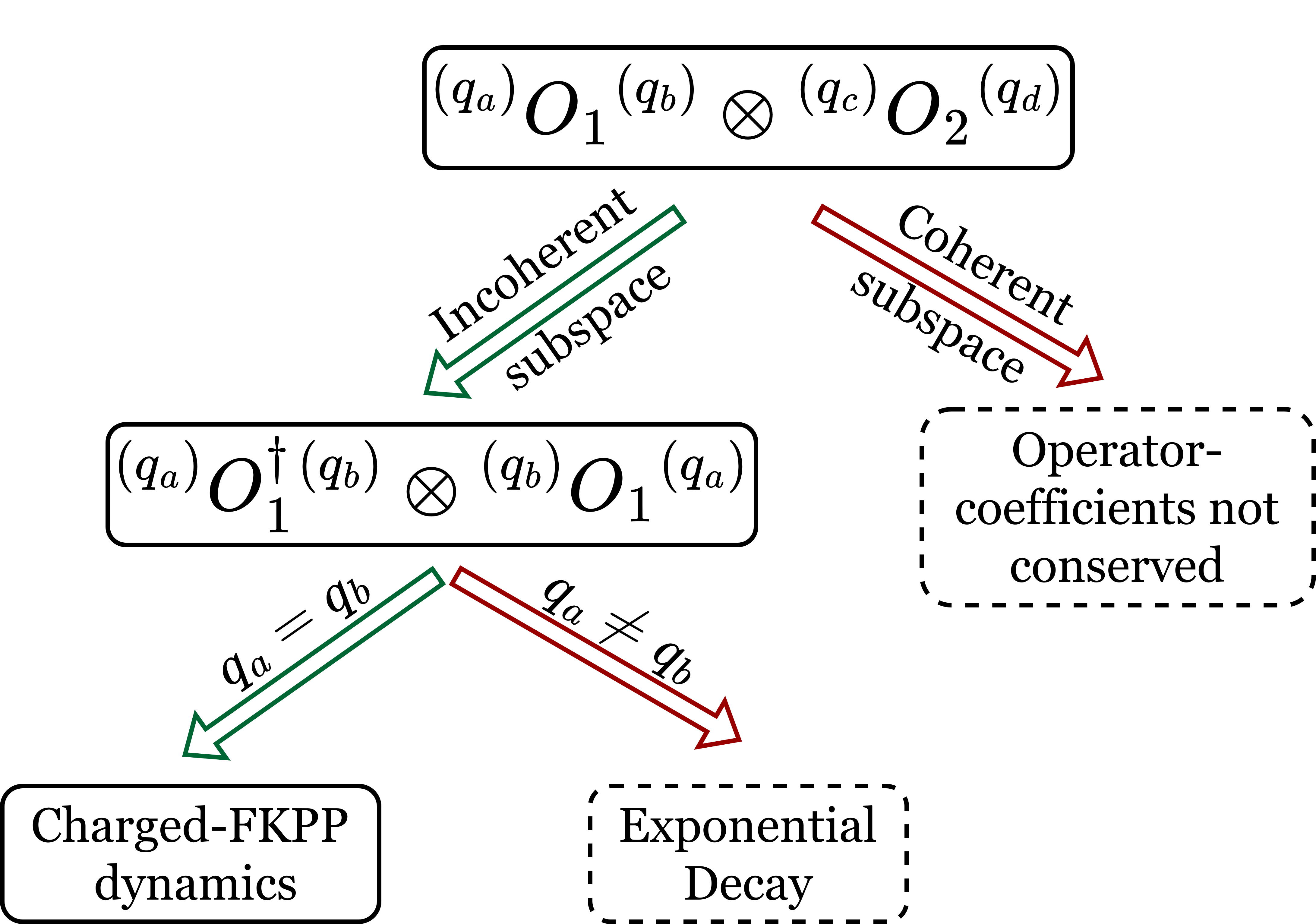}
\caption{Figure depicts the reduction of four charges involved in the computation of the OTOC, to a single charged mode. An operator-state of the form $O_1 \otimes O_2$, where each operator has two associated charges, has to be restricted to the incoherent subspace $(q_c = q_b \ ; \ q_d = q_a)$ in order for its coefficients to represent a conserved probability distribution. This reduces the four charges to two. Following this, the sector $q_a \neq q_b$ represents dynamics with pure exponential decay (as seen from Eq.~\eqref{Eq:left_right_FKPP}). Restricting to the single charge ($q_a = q_b$) sector enables the operator to show interesting dynamics such as Lyapunov growth and butterfly velocity.}
\label{Fig:charges_reduction}
\end{figure}

\subsection{Charge-dependent FKPP equation}
Adding interactions to the model, i.e. $H = H_{\text{inter}} + H_{\text{intra}}$, does not change the nature of the charge transport, as the added on-site interaction conserves charge on each site. However, this does change the behavior of the OTOC as the interaction term scrambles quantum information within each cluster. We write the equation obtained for $\xi$ in terms of the left($\rho_a$) and right($\rho_b$) charge densities as follows:
\bea
\label{Eq:left_right_FKPP}
&\partial_t \xi = g^2 \xi ( 2 \xi^2 - \rho_a(1-\rho_a) - \rho_b(1-\rho_b)) +  \partial_r^2 \xi \\
&\partial_t \rho_{a/b} = \partial_r^2 \rho_{a/b}
\eea
In general, the term proportional to $g^2$ hosts two roots, $\xi=0$ which is a stable solution and $\xi = \sqrt{(\rho_a(1-\rho_a)+\rho_b(1-\rho_b))/2}$ which is an unstable zero. However, one can obtain the physical range of $\xi$ by in turn determining the range of $n \rightarrow Nu$ through the constraints on the GT-pattern (Eq.~\eqref{Eq:GT-Patttern}) and also remembering the coordinate transformation in Eq.~(\ref{eq:cord_trans}):
\[0 \leq \xi \leq \text{min}(\sqrt{\rho_a (1-\rho_b)},\sqrt{(1-\rho_a)\rho_b})\]

Thus, when $\rho_a \neq \rho_b$, the physically maximal allowed value of $\xi$ is smaller than the unstable solution of the non-linear term in Eq.~(\ref{Eq:left_right_FKPP}). Therefore in this case, regardless of the initial profile of $\xi$, the non-linear term will be negative and the solution on every site will exponentially decay till it reaches the stable solution $\xi=0$. Physically, the stable root corresponds to the charge-resolved fully scrambled state $\sum_{\S} \ket{\S^{\dagger} \otimes \S}$. The unstable root  on the other hand is not a physical solution in general because it would physically correspond to the vanishing of the identity operator under the emergent Hamiltonian (Fig.~\ref{Fig:OTOC-Tensor}). For $\rho_a \neq \rho_b $, however, the charge sector does not host a component of the Identity (App.~\ref{App:Identity}). Therefore, the maximal allowed value of $\xi$ is not a solution and only the stable solution is physical. Hence no butterfly velocity or Lyapunov growth is observed when $\rho_a \neq \rho_b$. In terms of operators, the maximal value of $\xi$ in such a sector corresponds to a string with a large number of $\chi$ or $\chi^{\dagger}$, which means the strings are 'far away' from any component of the identity and therefore have very high energy (in terms of magnitude) with respect to the emergent Hamiltonian.

Now we will analyze the situation when $\rho_a = \rho_b$. Due to their equations of motion being structurally identical, $\partial_t (\rho_a - \rho_b) =0$ and we can conclude that starting with this condition will fix it for all time. For the charge sector where $\rho_a=\rho_b=\rho$, the coupled equations become:
\bea \label{eq:FKPP}
\partial_t \rho = \partial_r^2 \rho \; ; \; \partial_t \xi =2 g^2 \xi ( \xi^2 - \rho(1-\rho) ) + \partial_r^2 \xi.
\eea
This charge-dependent FKPP equation is one of the primary results of our work, which summarizes the effect of charge conservation on the OTOC. To briefly recap, we originally started with four variables, namely, three weights which represent charges on the OTOC-contour and one variable describing operator fluctuations within fixed-charge subspaces. One charge was fixed by requiring incoherence of the operators. Finally, we set the left and right charges to be equal to obtain the above equation describing the charge-dependent OTOC in terms of a single charged mode. This reduction in charges is depicted in Fig.~\ref{Fig:charges_reduction}. For the rest of this work, we will assume the use of the condition $\rho_a = \rho_b$.

\subsubsection{Constant charge density}

When $\rho$ is constant, the physically maximal allowed of $\xi$,  $\sqrt{\rho(1-\rho)}$, corresponds to the unstable solution of the FKPP-equation. The equation therefore supports travelling wave solutions $\xi(r,t) = f(r-v_B t)$ with minimal velocity $v_B$ when one starts with a source which is sufficiently well localized. Along with this, there is exponential growth of the field with Lyapunov exponent $\lambda_L$ ahead of the wave-front. The butterfly velocity and Lyapunov exponent can be computed as
\bea \label{eq:Lyap_Butterfly}
\lambda_L &= 8 g^2 \rho(1-\rho) \\
v_B &= 4 g \sqrt{ \rho(1-\rho) }
\eea
in terms of the charge density. Let's analyze these results in terms of operator-strings. The sector $\rho_a = \rho_b$ hosts null-eigenstates corresponding to the charge resolution of the identity $I_{q,N}$ (App.~\ref{App:Identity}), and starting from a state `near' $\xi= \sqrt{\rho(1-\rho)}$ corresponds to these operator-strings, which contain mostly $n$ and $\bn$ and very few $\chi^{\dagger}$ and $\chi$. These strings constitute the ground states of the emergent Hamiltonian, and perturbations to this ground state grow exponentially fast at early times and allow the solution to move with a well-defined velocity across the chain. The number of $n$'s in the said string determines the value $\rho$, and therefore also the speed of the butterfly velocity and the rate of exponential growth. On the other hand, the stable solution $\xi=0$ represents the charge-resolved, fully scrambled steady-state, i.e.\ $\sum_{\S} \ket{\S^{\dagger} \otimes \S}$ restricted to the region $\rho_a=\rho_b=\rho$.

\subsubsection{Charge-transport and late-time behavior}
\begin{figure}
\includegraphics[scale=0.5]{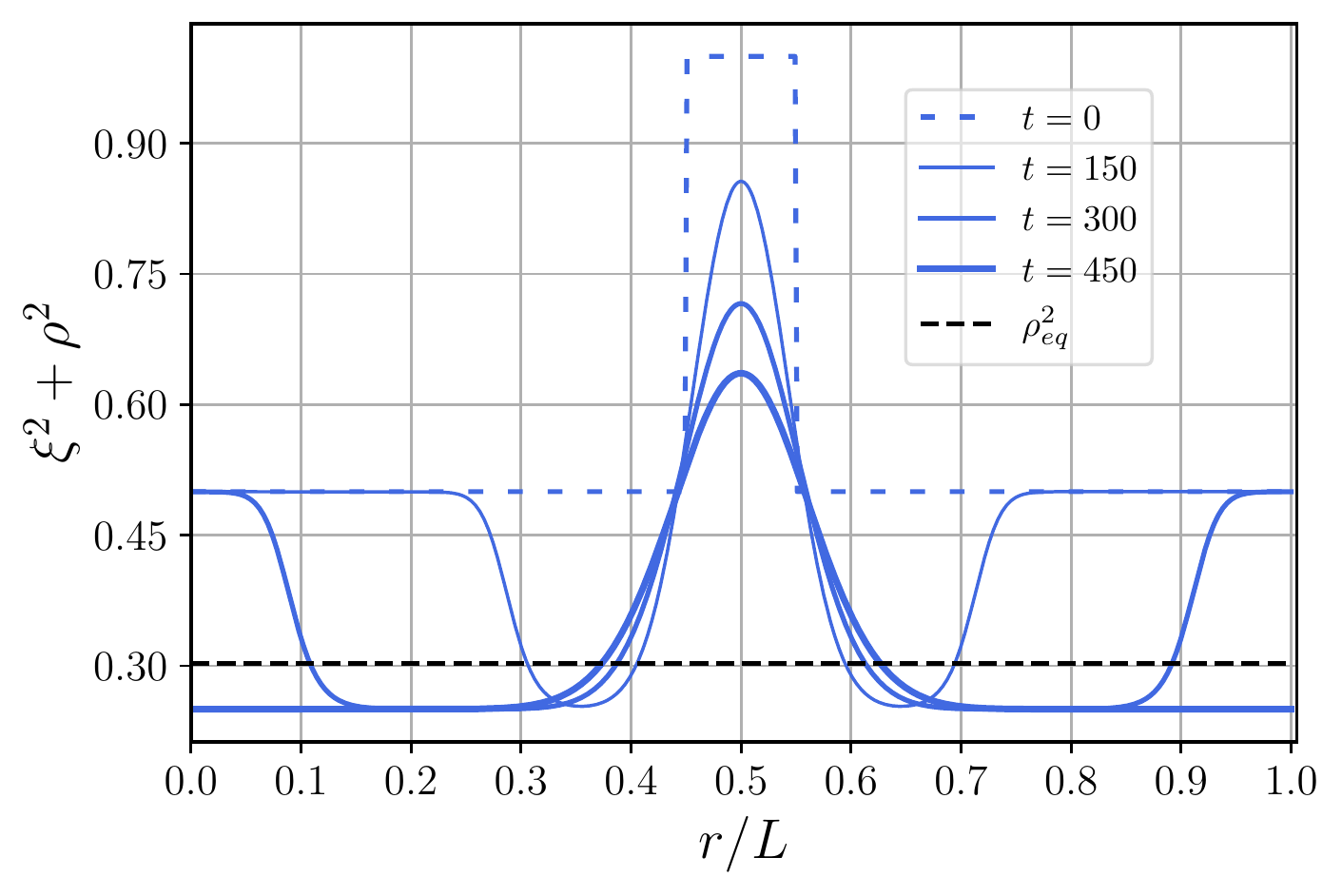}
\caption{The function $\xi^2 + \rho^2$, which corresponds to the OTOC when the probing operator is $V=n_{r_0}$, is plotted at equal time-intervals. In this scenario, we start with $\rho=1/2$ at every site except near the center where we place $\rho =1$ for sites $0.45 \leq r/L \leq 0.55$. The variable $\xi$ starts with its maximal value given by $\xi(t=0) = \sqrt{\rho(1-\rho)}$ on every site. We also impose open boundary conditions and set $2 g^2=0.1$. The two equations in Eq.~\eqref{eq:FKPP} are then simulated and we observe that while a part of the function corresponding to the non-conserved $\xi$ expands ballistically, the conserved part $\rho$ spreads out diffusively near the center. This slower mode controls the late-time behavior after the ballistic part has travelled across the chain, as can be seen from the $t=450$ time-step. The OTOC relaxes diffusively to the equilibrium value given by $\rho^2_{eq} \sim 0.3$ (the dashed black line) at very late times.}
\label{Fig:conserved_operator}
\end{figure}

So far we have considered the operator $\chi_{r_0}$, which means that the OTOC is given by the variable $\F=\xi_{r_0}(t)$. This ensures that the OTOC decays to 0, which is related to the fact that the operator $\chi_{r_0}$ has zero overlap with the charge. In this case, the late-time behavior of the OTOC is always determined by the non-linear term in the FKPP equation, which is represented by an exponential decay. 

In contrast, fixing the probe operator in the OTOC to be $V = n_{r_{0}}$ modifies the output state and makes it so that the OTOC is given by $\F = \xi_{r_0}^2(t) + \rho_{r_0}^2(t)$. This function is plotted in Fig.~\ref{Fig:conserved_operator}, where we start with an initial charge-profile which is non-uniform. We observe that there is a mode, which does not have overlap with the charge, that travels ballistically as well a charged mode which is 'left-behind' near the center \cite{Khemani_2018, Rakovszky_2018}. Since the equations of motion (Eq.~\eqref{eq:FKPP}) remain unchanged, the OTOC does not decay to 0, but instead at late-times is determined by the equilibrium value of the charge-distribution. In such a scenario, the late-time behavior of the OTOC is controlled by the nature of the charge-transport, since $\rho$ decays slower than $\xi$ and is therefore the dominant mode at late times. As an example, generalizing the model in this work to $d$ spatial dimensions will cause the charge to decay diffusively with the power-law $\sim t^{-d/2}$, which will therefore be the late-time behavior of the OTOC as well. This explains the observed behavior of the OTOC studied in similar models with U(1) conservation \cite{cheng2021scrambling, Khemani_2018, Rakovszky_2018}.

\section{A case study: operator dynamics in a domain wall density background}
\label{sec:Domain_wall}
In this section we will discuss operator dynamics in the case where the charge-background starts in a domain-wall configuration. Depending on the initial left and right density, i.e.\ $\rho_L$ and $\rho_R$, we gain different insights into how the charge dynamics can affect operator growth. We will assume that $\xi$ always starts with its maximal initial value $\sqrt{\rho(1-\rho)}$.

\begin{figure}
\includegraphics[scale=0.5]{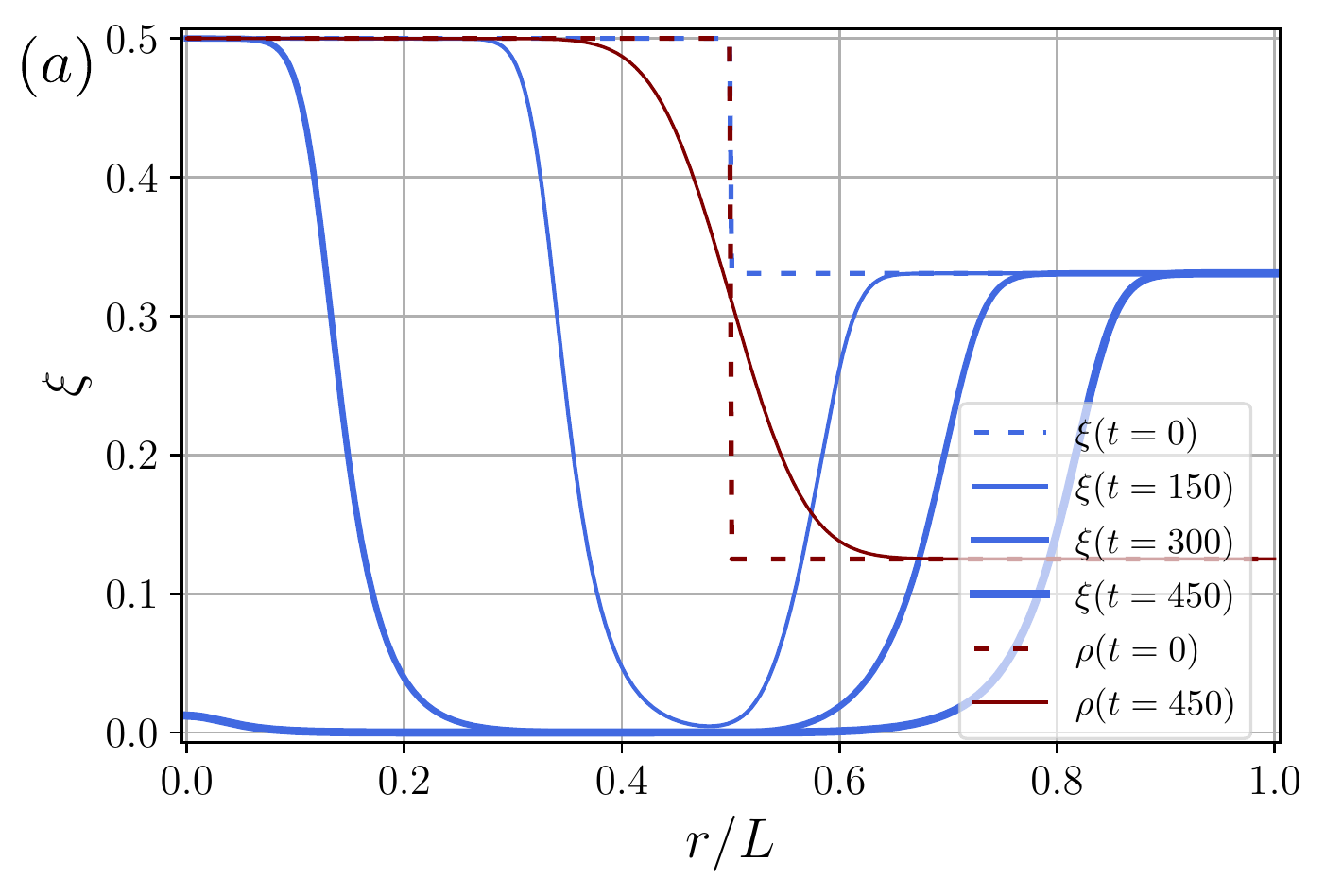}
\includegraphics[scale=0.5]{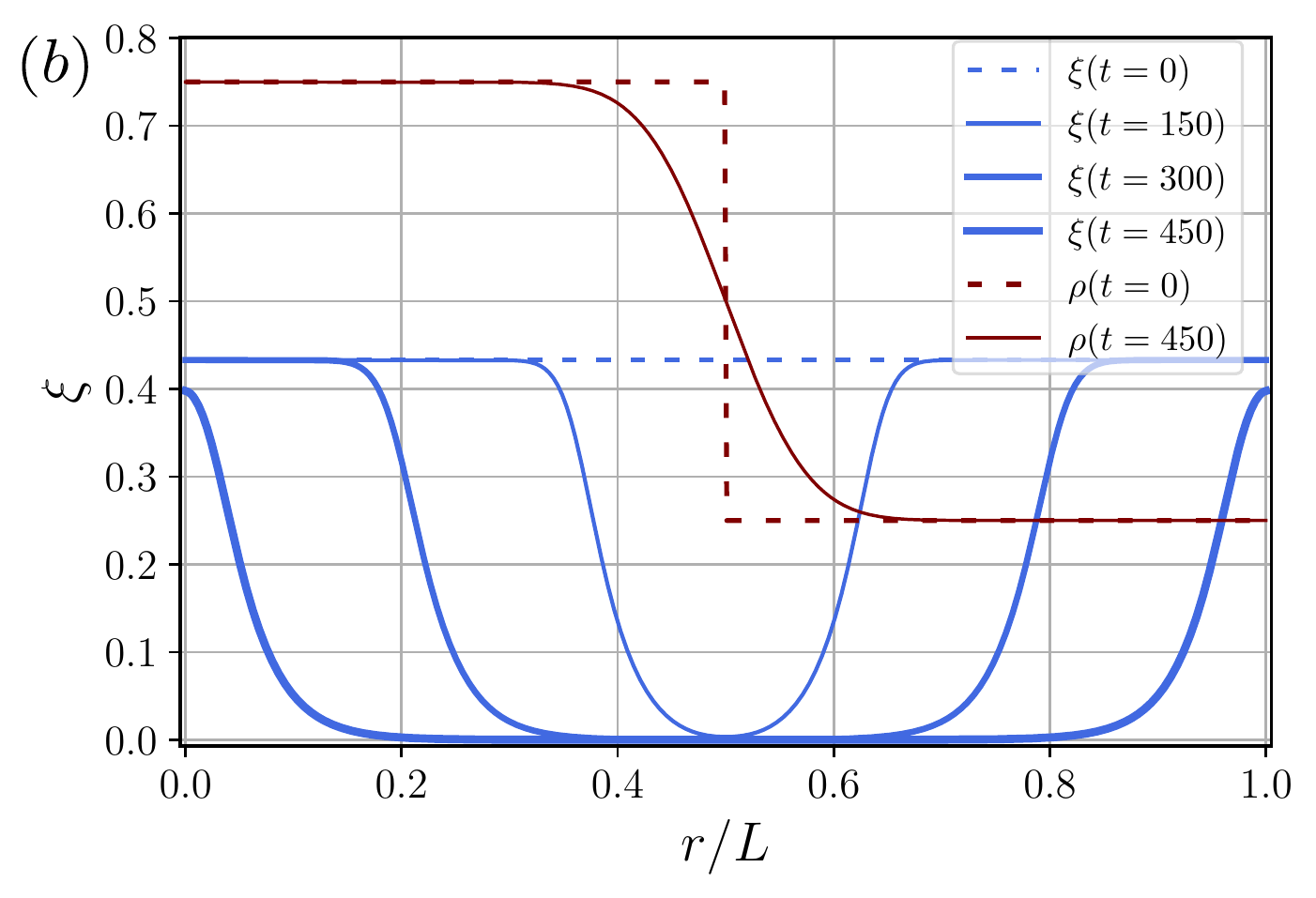}
\caption{The solutions of Eq.~\eqref{eq:FKPP} are plotted for two different charge profiles. We start with the maximally allowed value of $\xi(t=0) = \sqrt{\rho(1-\rho)}$ and impose open boundary conditions in both cases, along with the condition $2g^2=0.1$. In (a), a charge domain wall is used with density $1/2$ on the left and $1/8$ on the right. This causes the operator to acquire an asymmetric velocity which is evident when observing the wavefront at equal time-intervals. In (b), the left and right charge densities are $3/4$ and $1/4$ respectively, causing the operator to have a flat profile which is static by itself. However, the melting of the domain-wall at early times forces the operator to have dynamics, even though it has zero overlap with the charge.}
\label{Fig:charge_scrambling}
\end{figure}

\subsection{Asymmetric butterfly velocity}

The first case we consider involves the left-half of the chain initialized in density-profile $\rho_L=1/2$ and the right-half being in $\rho_R = 1/8$. The operator starts with its maximal value $\sqrt{\rho(1-\rho)}$ on every site. In this scenario, both the varying $\rho$ and $\xi$ are perturbed by the diffusion term at $t=0$, and the non-linear term for $\xi$ then carries the perturbation along the chain with a butterfly velocity. Due to the nature of the domain wall, the left and right halves of the operator can travel with different speeds. This phenomenon is depicted in Fig.~\ref{Fig:charge_scrambling}(a), where the wavefront plotted at equal time intervals clearly demonstrates that the operator moves faster through the left half of the chain, as expected from Eq.~\eqref{eq:Lyap_Butterfly}. It is also clear that the domain wall melts at a rate much slower than the evolution of the operator, which makes the dynamics of the charge largely inconsequential to the dynamics of the operator $\chi$, in this scenario.

\subsection{Conserved vs.\ non-conserved operator}

If one starts with a completely filled density on one half $(\rho_L = 1)$ while the other remains completely unoccupied $(\rho_R = 0)$, the variable $\xi$ can only take one value, i.e.\ 0, at every site. Since this is a stable solution of the FKPP-equation, the OTOC involving the operator $\chi$ remains completely static, even though the charge and therefore the on-site scrambled-state changes with time. However, if one uses a conserved operator, such as $n$, the OTOC is captured by the variable $\xi^2 + \rho^2$. Since the charge moves diffusively as the domain wall melts, it causes the OTOC to have diffusive behavior at all times. Hence, in this situation, the OTOC involving an operator such as $\chi$ remains perfectly static, whereas the OTOC for a conserved operator such as $n$ shows diffusive behavior.

\subsection{Charge dynamics influences the dynamics of non-conserved operators}
\label{subsec:Kickstart_operator}
Let's consider a scenario where the left and right densities are chosen to be $\rho_L=3/4$ and $\rho_R=1/4$ respectively (Fig.~\ref{Fig:charge_scrambling}(b)). This makes it so that the value of $\rho(1-\rho)$ is equal on both sides of the chain, even though the density has different values. Thus, at $t=0$, the operator-profile which starts at its maximal value, is flat and static, although the charge is not. This case is different from the previous one because the operator dynamics are kick-started purely due to the charge dynamics, where at early times the diffusion for the charge causes the domain wall to melt and therefore forces the operator to start moving as well. In fact, this can be generalized further. When starting with some specific operator-profile $\xi(r,t=0) = f(r)$, there are two available solutions for the corresponding charge density since $\xi = \sqrt{\rho(1-\rho)}$. Each solution for $\rho$ individually gives the same behavior for $\xi$, however starting from different solutions on the left and right half of the chain creates a perturbation in the center as the charge moves to bridge the gap between the two solutions. This is especially interesting because usually one expects the transport to only be relevant when conserved operators such as $n$ are involved, which have overlap with the charge. However, even though the operator $\chi$ does not have overlap with the charge, the charge dynamics in the background forces the Hilbert-space dimension and therefore the maximum allowed value of $\xi$ to change in the middle of the chain. This causes a local perturbation that then travels ballistically across the chain.

\section{Conclusion and Discussions}
\label{sec:conclusions}
In this work we utilize the symmetry structures present in Brownian SYK models, which emerge after the disorder averaging procedure, to study the relation between charge-transport and operator growth. We measure operator growth and scrambling by computing the OTOC in the complex Brownian SYK chain with U(1) symmetry. For this calculation, the model maps to an SU(4) spin chain which evolves in imaginary time and conserves weights. The computation of the OTOC involves four time-contours, and therefore four conserved charges as well. In our formalism, we show that the dynamics of these charges can be encoded in terms of the evolution of the conserved weights. Moreover, the evolution of operator-states is represented by transitions between states in the SU(4) algebra. 

Tracking these transitions reveals distinct features of operator dynamics in many-body models with symmetries. While usual Brownian/random models can be modeled by a classical stochastic process, we show that the U(1) symmetry introduces quantum-coherence at the operator level, which only allows for a classical description in a subspace of the operator-space. In the infinite-$N$ limit, the model therefore is described by a `restricted' Fokker-Planck equation, which represents a probability distribution only in the sector where the operator-states have no coherence.

We utilize the mapping to solve the OTOC and charge-transport exactly at finite-$N$, for the non-interacting model. For the model with both transport and scrambling, we compute the OTOC involving simple operators, which can be shown to lie in the incoherent subspace, in the large-$N$ limit. This enables us to derive an FKPP-equation governing the evolution of the operator coupled to the charge which is valid for all charge-density profiles and at all time-scales. For constant charge-density background, the coupled FKPP equation provides us with the Lyapunov exponent and the butterfly velocity as a function of the charge density. Using these equations we also explain known results for charged-operator dynamics, such as the diffusive late-time behavior, and explore new regimes of varying charge-density backgrounds. When starting with different domain-wall solutions, we obtain novel solutions for operator dynamics, including one where the dynamics even for a non-conserved operator such as $\chi$ are kick-started purely due to the charge dynamics. Next, we share some additional insights involving operator growth in the presence of symmetries.

In this work, we have considered a model which moves the charge and scrambles quantum information simultaneously. However, it is possible to decouple the movement of the charge from the spatial movement of the operator, by considering a Brownian Hamiltonian like:
\bea
\label{eq:interacting_charge_conserve}
H = \sum_{i,j,k,l,r} J_{i,j,k,l,r}(t) \chi^{\dagger}_{i,r} \chi_{j,r} \chi^{\dagger}_{k,r+1} \chi_{l,r+1} + \text{h.c.}
\eea
This model is interacting, yet does not allow the transfer of charge as it conserves charge on each site $r$. It can be solved in the large-$N$ limit as well, where the equation for $\xi$ (the OTOC) has a stable zero at $\xi=0$ and a transport term for $\xi$, but not the charge. This can be contrasted with the case in Sec.~\ref{sec:Domain_wall}.2, in which case the charge transport was required to kickstart the operator dynamics. Thus, if one were to consider the model in Eq.~\eqref{eq:interacting_charge_conserve} instead, the OTOC in that scenario would remain completely static. 

In contrast with this, if one considers a model which conserves energy, it does not seem possible to construct scenarios which scramble information yet do not simultaneously transport energy. This marks a stark difference between a gauge symmetry, such as charge conservation, and energy conservation. 

\section{Acknowledgement}
We thank Brian Swingle and Christopher M. Langlett for helpful discussions and comments on the manuscript. L. A. thanks Naven Narayanan for pointing out the connection between bacterial population dynamics and the operator growth model obtained in this work.

Research at Perimeter Institute is supported in part by the Government of Canada through the Department of Innovation, Science and Economic Development Canada and by the Province of Ontario through the Ministry of Colleges and Universities.

\bibliography{ref}

\begin{thebibliography}{75}%
\makeatletter
\providecommand \@ifxundefined [1]{%
 \@ifx{#1\undefined}
}%
\providecommand \@ifnum [1]{%
 \ifnum #1\expandafter \@firstoftwo
 \else \expandafter \@secondoftwo
 \fi
}%
\providecommand \@ifx [1]{%
 \ifx #1\expandafter \@firstoftwo
 \else \expandafter \@secondoftwo
 \fi
}%
\providecommand \natexlab [1]{#1}%
\providecommand \enquote  [1]{``#1''}%
\providecommand \bibnamefont  [1]{#1}%
\providecommand \bibfnamefont [1]{#1}%
\providecommand \citenamefont [1]{#1}%
\providecommand \href@noop [0]{\@secondoftwo}%
\providecommand \href [0]{\begingroup \@sanitize@url \@href}%
\providecommand \@href[1]{\@@startlink{#1}\@@href}%
\providecommand \@@href[1]{\endgroup#1\@@endlink}%
\providecommand \@sanitize@url [0]{\catcode `\\12\catcode `\$12\catcode
  `\&12\catcode `\#12\catcode `\^12\catcode `\_12\catcode `\%12\relax}%
\providecommand \@@startlink[1]{}%
\providecommand \@@endlink[0]{}%
\providecommand \url  [0]{\begingroup\@sanitize@url \@url }%
\providecommand \@url [1]{\endgroup\@href {#1}{\urlprefix }}%
\providecommand \urlprefix  [0]{URL }%
\providecommand \Eprint [0]{\href }%
\providecommand \doibase [0]{https://doi.org/}%
\providecommand \selectlanguage [0]{\@gobble}%
\providecommand \bibinfo  [0]{\@secondoftwo}%
\providecommand \bibfield  [0]{\@secondoftwo}%
\providecommand \translation [1]{[#1]}%
\providecommand \BibitemOpen [0]{}%
\providecommand \bibitemStop [0]{}%
\providecommand \bibitemNoStop [0]{.\EOS\space}%
\providecommand \EOS [0]{\spacefactor3000\relax}%
\providecommand \BibitemShut  [1]{\csname bibitem#1\endcsname}%
\let\auto@bib@innerbib\@empty
\bibitem [{\citenamefont {Deutsch}(1991)}]{Deutsch1991}%
  \BibitemOpen
  \bibfield  {author} {\bibinfo {author} {\bibfnamefont {J.~M.}\ \bibnamefont
  {Deutsch}},\ }\bibfield  {title} {\bibinfo {title} {{Quantum statistical
  mechanics in a closed system}},\ }\href
  {https://doi.org/10.1103/PhysRevA.43.2046} {\bibfield  {journal} {\bibinfo
  {journal} {Phys. Rev. A}\ }\textbf {\bibinfo {volume} {43}},\ \bibinfo
  {pages} {2046} (\bibinfo {year} {1991})}\BibitemShut {NoStop}%
\bibitem [{\citenamefont {Srednicki}(1994)}]{Srednicki1994}%
  \BibitemOpen
  \bibfield  {author} {\bibinfo {author} {\bibfnamefont {M.}~\bibnamefont
  {Srednicki}},\ }\bibfield  {title} {\bibinfo {title} {{Chaos and quantum
  thermalization}},\ }\href {https://doi.org/10.1103/PhysRevE.50.888}
  {\bibfield  {journal} {\bibinfo  {journal} {Phys. Rev. E}\ }\textbf {\bibinfo
  {volume} {50}},\ \bibinfo {pages} {888} (\bibinfo {year} {1994})}\BibitemShut
  {NoStop}%
\bibitem [{\citenamefont {Rigol}\ \emph {et~al.}(2008)\citenamefont {Rigol},
  \citenamefont {Dunjko},\ and\ \citenamefont
  {Olshanii}}]{rigol2008thermalization}%
  \BibitemOpen
  \bibfield  {author} {\bibinfo {author} {\bibfnamefont {M.}~\bibnamefont
  {Rigol}}, \bibinfo {author} {\bibfnamefont {V.}~\bibnamefont {Dunjko}},\ and\
  \bibinfo {author} {\bibfnamefont {M.}~\bibnamefont {Olshanii}},\ }\bibfield
  {title} {\bibinfo {title} {Thermalization and its mechanism for generic
  isolated quantum systems},\ }\href {https://doi.org/10.1038/nature06838}
  {\bibfield  {journal} {\bibinfo  {journal} {Nature}\ }\textbf {\bibinfo
  {volume} {452}},\ \bibinfo {pages} {854} (\bibinfo {year}
  {2008})}\BibitemShut {NoStop}%
\bibitem [{\citenamefont {Polkovnikov}\ \emph {et~al.}(2011)\citenamefont
  {Polkovnikov}, \citenamefont {Sengupta}, \citenamefont {Silva},\ and\
  \citenamefont {Vengalattore}}]{Polkovnikov2011}%
  \BibitemOpen
  \bibfield  {author} {\bibinfo {author} {\bibfnamefont {A.}~\bibnamefont
  {Polkovnikov}}, \bibinfo {author} {\bibfnamefont {K.}~\bibnamefont
  {Sengupta}}, \bibinfo {author} {\bibfnamefont {A.}~\bibnamefont {Silva}},\
  and\ \bibinfo {author} {\bibfnamefont {M.}~\bibnamefont {Vengalattore}},\
  }\bibfield  {title} {\bibinfo {title} {Colloquium: Nonequilibrium dynamics of
  closed interacting quantum systems},\ }\href
  {https://doi.org/10.1103/RevModPhys.83.863} {\bibfield  {journal} {\bibinfo
  {journal} {Rev. Mod. Phys.}\ }\textbf {\bibinfo {volume} {83}},\ \bibinfo
  {pages} {863} (\bibinfo {year} {2011})}\BibitemShut {NoStop}%
\bibitem [{\citenamefont {Nahum}\ \emph {et~al.}(2018)\citenamefont {Nahum},
  \citenamefont {Vijay},\ and\ \citenamefont {Haah}}]{Nahum_2018}%
  \BibitemOpen
  \bibfield  {author} {\bibinfo {author} {\bibfnamefont {A.}~\bibnamefont
  {Nahum}}, \bibinfo {author} {\bibfnamefont {S.}~\bibnamefont {Vijay}},\ and\
  \bibinfo {author} {\bibfnamefont {J.}~\bibnamefont {Haah}},\ }\bibfield
  {title} {\bibinfo {title} {Operator spreading in random unitary circuits},\
  }\href {http://dx.doi.org/10.1103/PhysRevX.8.021014} {\bibfield  {journal}
  {\bibinfo  {journal} {Phys. Rev. X}\ }\textbf {\bibinfo {volume} {8}},\
  \bibinfo {pages} {021014} (\bibinfo {year} {2018})}\BibitemShut {NoStop}%
\bibitem [{\citenamefont {Von~Keyserlingk}\ \emph {et~al.}(2018)\citenamefont
  {Von~Keyserlingk}, \citenamefont {Rakovszky}, \citenamefont {Pollmann},\ and\
  \citenamefont {Sondhi}}]{von2018operator}%
  \BibitemOpen
  \bibfield  {author} {\bibinfo {author} {\bibfnamefont {C.}~\bibnamefont
  {Von~Keyserlingk}}, \bibinfo {author} {\bibfnamefont {T.}~\bibnamefont
  {Rakovszky}}, \bibinfo {author} {\bibfnamefont {F.}~\bibnamefont
  {Pollmann}},\ and\ \bibinfo {author} {\bibfnamefont {S.~L.}\ \bibnamefont
  {Sondhi}},\ }\bibfield  {title} {\bibinfo {title} {Operator hydrodynamics,
  otocs, and entanglement growth in systems without conservation laws},\ }\href
  {https://link.aps.org/doi/10.1103/PhysRevX.8.021013} {\bibfield  {journal}
  {\bibinfo  {journal} {Phys. Rev. X}\ }\textbf {\bibinfo {volume} {8}},\
  \bibinfo {pages} {021013} (\bibinfo {year} {2018})}\BibitemShut {NoStop}%
\bibitem [{\citenamefont {Parker}\ \emph {et~al.}(2019)\citenamefont {Parker},
  \citenamefont {Cao}, \citenamefont {Avdoshkin}, \citenamefont {Scaffidi},\
  and\ \citenamefont {Altman}}]{parker2019universal}%
  \BibitemOpen
  \bibfield  {author} {\bibinfo {author} {\bibfnamefont {D.~E.}\ \bibnamefont
  {Parker}}, \bibinfo {author} {\bibfnamefont {X.}~\bibnamefont {Cao}},
  \bibinfo {author} {\bibfnamefont {A.}~\bibnamefont {Avdoshkin}}, \bibinfo
  {author} {\bibfnamefont {T.}~\bibnamefont {Scaffidi}},\ and\ \bibinfo
  {author} {\bibfnamefont {E.}~\bibnamefont {Altman}},\ }\bibfield  {title}
  {\bibinfo {title} {A universal operator growth hypothesis},\ }\href
  {https://link.aps.org/doi/10.1103/PhysRevX.9.041017} {\bibfield  {journal}
  {\bibinfo  {journal} {Phys. Rev. X}\ }\textbf {\bibinfo {volume} {9}},\
  \bibinfo {pages} {041017} (\bibinfo {year} {2019})}\BibitemShut {NoStop}%
\bibitem [{\citenamefont {Hosur}\ \emph {et~al.}(2016)\citenamefont {Hosur},
  \citenamefont {Qi}, \citenamefont {Roberts},\ and\ \citenamefont
  {Yoshida}}]{Hosur_2016}%
  \BibitemOpen
  \bibfield  {author} {\bibinfo {author} {\bibfnamefont {P.}~\bibnamefont
  {Hosur}}, \bibinfo {author} {\bibfnamefont {X.-L.}\ \bibnamefont {Qi}},
  \bibinfo {author} {\bibfnamefont {D.~A.}\ \bibnamefont {Roberts}},\ and\
  \bibinfo {author} {\bibfnamefont {B.}~\bibnamefont {Yoshida}},\ }\bibfield
  {title} {\bibinfo {title} {Chaos in quantum channels},\ }\href
  {http://doi.org/10.1007/JHEP02(2016)004} {\bibfield  {journal} {\bibinfo
  {journal} {J. High Energy Phys.}\ }\textbf {\bibinfo {volume} {2016}}\bibinfo
   {number} { (02)},\ \bibinfo {pages} {004}}\BibitemShut {NoStop}%
\bibitem [{\citenamefont {Keselman}\ \emph {et~al.}(2021)\citenamefont
  {Keselman}, \citenamefont {Nie},\ and\ \citenamefont
  {Berg}}]{keselman2021scrambling}%
  \BibitemOpen
\bibfield  {number} {  }\bibfield  {author} {\bibinfo {author} {\bibfnamefont
  {A.}~\bibnamefont {Keselman}}, \bibinfo {author} {\bibfnamefont
  {L.}~\bibnamefont {Nie}},\ and\ \bibinfo {author} {\bibfnamefont
  {E.}~\bibnamefont {Berg}},\ }\bibfield  {title} {\bibinfo {title} {Scrambling
  and lyapunov exponent in spatially extended systems},\ }\href
  {https://link.aps.org/doi/10.1103/PhysRevB.103.L121111} {\bibfield  {journal}
  {\bibinfo  {journal} {Phys. Rev. B}\ }\textbf {\bibinfo {volume} {103}},\
  \bibinfo {pages} {L121111} (\bibinfo {year} {2021})}\BibitemShut {NoStop}%
\bibitem [{\citenamefont {Knap}(2018)}]{knap2018entanglement}%
  \BibitemOpen
  \bibfield  {author} {\bibinfo {author} {\bibfnamefont {M.}~\bibnamefont
  {Knap}},\ }\bibfield  {title} {\bibinfo {title} {Entanglement production and
  information scrambling in a noisy spin system},\ }\href
  {https://link.aps.org/doi/10.1103/PhysRevB.98.184416} {\bibfield  {journal}
  {\bibinfo  {journal} {Phys. Rev. B}\ }\textbf {\bibinfo {volume} {98}},\
  \bibinfo {pages} {184416} (\bibinfo {year} {2018})}\BibitemShut {NoStop}%
\bibitem [{\citenamefont {Sekino}\ and\ \citenamefont
  {Susskind}()}]{Sekino_2008}%
  \BibitemOpen
  \bibfield  {author} {\bibinfo {author} {\bibfnamefont {Y.}~\bibnamefont
  {Sekino}}\ and\ \bibinfo {author} {\bibfnamefont {L.}~\bibnamefont
  {Susskind}},\ }\bibfield  {title} {\bibinfo {title} {Fast scramblers},\
  }\href {https://doi.org/10.1088/1126-6708/2008/10/065} {\bibfield  {journal}
  {\bibinfo  {journal} {J. High Energy Phys.}\ }\textbf {\bibinfo {volume}
  {2008}}\bibinfo  {number} { (10)},\ \bibinfo {pages} {065}}\BibitemShut
  {NoStop}%
\bibitem [{\citenamefont {Hayden}\ and\ \citenamefont
  {Preskill}(2007)}]{Hayden_2007}%
  \BibitemOpen
\bibfield  {number} {  }\bibfield  {author} {\bibinfo {author} {\bibfnamefont
  {P.}~\bibnamefont {Hayden}}\ and\ \bibinfo {author} {\bibfnamefont
  {J.}~\bibnamefont {Preskill}},\ }\bibfield  {title} {\bibinfo {title} {Black
  holes as mirrors: quantum information in random subsystems},\ }\href
  {https://doi.org/10.1088/1126-6708/2007/09/120} {\bibfield  {journal}
  {\bibinfo  {journal} {J. High Energy Phys.}\ }\textbf {\bibinfo {volume}
  {2007}}\bibinfo  {number} { (09)},\ \bibinfo {pages} {120}}\BibitemShut
  {NoStop}%
\bibitem [{\citenamefont {Shenker}\ and\ \citenamefont
  {Stanford}(2015)}]{shenker2015stringy}%
  \BibitemOpen
\bibfield  {number} {  }\bibfield  {author} {\bibinfo {author} {\bibfnamefont
  {S.~H.}\ \bibnamefont {Shenker}}\ and\ \bibinfo {author} {\bibfnamefont
  {D.}~\bibnamefont {Stanford}},\ }\bibfield  {title} {\bibinfo {title}
  {Stringy effects in scrambling},\ }\href
  {https://doi.org/10.1007/JHEP05(2015)132} {\bibfield  {journal} {\bibinfo
  {journal} {J. High Energy Phys.}\ }\textbf {\bibinfo {volume} {2015}}\bibinfo
   {number} { (05)},\ \bibinfo {pages} {132}}\BibitemShut {NoStop}%
\bibitem [{\citenamefont {Maldacena}\ \emph {et~al.}(2016)\citenamefont
  {Maldacena}, \citenamefont {Shenker},\ and\ \citenamefont
  {Stanford}}]{Maldacena_2016}%
  \BibitemOpen
\bibfield  {number} {  }\bibfield  {author} {\bibinfo {author} {\bibfnamefont
  {J.}~\bibnamefont {Maldacena}}, \bibinfo {author} {\bibfnamefont {S.~H.}\
  \bibnamefont {Shenker}},\ and\ \bibinfo {author} {\bibfnamefont
  {D.}~\bibnamefont {Stanford}},\ }\bibfield  {title} {\bibinfo {title} {A
  bound on chaos},\ }\href {https://doi.org/10.1007/JHEP08(2016)106} {\bibfield
   {journal} {\bibinfo  {journal} {J. High Energy Phys.}\ }\textbf {\bibinfo
  {volume} {2016}}\bibinfo  {number} { (08)},\ \bibinfo {pages}
  {106}}\BibitemShut {NoStop}%
\bibitem [{\citenamefont {Shenker}\ and\ \citenamefont
  {Stanford}(2014)}]{shenker2014black}%
  \BibitemOpen
\bibfield  {number} {  }\bibfield  {author} {\bibinfo {author} {\bibfnamefont
  {S.~H.}\ \bibnamefont {Shenker}}\ and\ \bibinfo {author} {\bibfnamefont
  {D.}~\bibnamefont {Stanford}},\ }\bibfield  {title} {\bibinfo {title} {Black
  holes and the butterfly effect},\ }\href
  {https://doi.org/10.1007/JHEP03(2014)067} {\bibfield  {journal} {\bibinfo
  {journal} {J. High Energy Phys.}\ }\textbf {\bibinfo {volume} {2014}}\bibinfo
   {number} { (03)},\ \bibinfo {pages} {067}}\BibitemShut {NoStop}%
\bibitem [{\citenamefont {Lashkari}\ \emph {et~al.}(2013)\citenamefont
  {Lashkari}, \citenamefont {Stanford}, \citenamefont {Hastings}, \citenamefont
  {Osborne},\ and\ \citenamefont {Hayden}}]{Lashkari_2013}%
  \BibitemOpen
\bibfield  {number} {  }\bibfield  {author} {\bibinfo {author} {\bibfnamefont
  {N.}~\bibnamefont {Lashkari}}, \bibinfo {author} {\bibfnamefont
  {D.}~\bibnamefont {Stanford}}, \bibinfo {author} {\bibfnamefont
  {M.}~\bibnamefont {Hastings}}, \bibinfo {author} {\bibfnamefont
  {T.}~\bibnamefont {Osborne}},\ and\ \bibinfo {author} {\bibfnamefont
  {P.}~\bibnamefont {Hayden}},\ }\bibfield  {title} {\bibinfo {title} {Towards
  the fast scrambling conjecture},\ }\href
  {http://dx.doi.org/10.1007/JHEP04(2013)022} {\bibfield  {journal} {\bibinfo
  {journal} {J. High Energy Phys.}\ }\textbf {\bibinfo {volume} {2013}}\bibinfo
   {number} { (04)},\ \bibinfo {pages} {022}}\BibitemShut {NoStop}%
\bibitem [{\citenamefont {Polchinski}\ and\ \citenamefont
  {Rosenhaus}(2016)}]{Polchinski2016}%
  \BibitemOpen
\bibfield  {number} {  }\bibfield  {author} {\bibinfo {author} {\bibfnamefont
  {J.}~\bibnamefont {Polchinski}}\ and\ \bibinfo {author} {\bibfnamefont
  {V.}~\bibnamefont {Rosenhaus}},\ }\bibfield  {title} {\bibinfo {title} {The
  spectrum in the {Sachdev-Ye-Kitaev model}},\ }\href
  {http://dx.doi.org/10.1007/JHEP04(2016)001} {\bibfield  {journal} {\bibinfo
  {journal} {J. High Energy Phys.}\ }\textbf {\bibinfo {volume} {2016}}\bibinfo
   {number} { (04)},\ \bibinfo {pages} {001}}\BibitemShut {NoStop}%
\bibitem [{\citenamefont {Maldacena}\ and\ \citenamefont
  {Stanford}(2016)}]{Maldacena_2016_SYK}%
  \BibitemOpen
\bibfield  {number} {  }\bibfield  {author} {\bibinfo {author} {\bibfnamefont
  {J.}~\bibnamefont {Maldacena}}\ and\ \bibinfo {author} {\bibfnamefont
  {D.}~\bibnamefont {Stanford}},\ }\bibfield  {title} {\bibinfo {title}
  {Remarks on the {Sachdev-Ye-Kitaev} model},\ }\href
  {http://dx.doi.org/10.1103/PhysRevD.94.106002} {\bibfield  {journal}
  {\bibinfo  {journal} {Phys. Rev. D}\ }\textbf {\bibinfo {volume} {94}},\
  \bibinfo {pages} {106002} (\bibinfo {year} {2016})}\BibitemShut {NoStop}%
\bibitem [{\citenamefont {Gross}\ and\ \citenamefont
  {Rosenhaus}(2017)}]{Gross_2017}%
  \BibitemOpen
  \bibfield  {author} {\bibinfo {author} {\bibfnamefont {D.~J.}\ \bibnamefont
  {Gross}}\ and\ \bibinfo {author} {\bibfnamefont {V.}~\bibnamefont
  {Rosenhaus}},\ }\bibfield  {title} {\bibinfo {title} {A generalization of
  sachdev-ye-kitaev},\ }\href {http://doi.org/10.1007/JHEP02(2017)093}
  {\bibfield  {journal} {\bibinfo  {journal} {J. High Energy Phys.}\ }\textbf
  {\bibinfo {volume} {2017}}\bibinfo  {number} { (02)}}\BibitemShut {NoStop}%
\bibitem [{\citenamefont {Gu}\ \emph {et~al.}(2020)\citenamefont {Gu},
  \citenamefont {Kitaev}, \citenamefont {Sachdev},\ and\ \citenamefont
  {Tarnopolsky}}]{Gu_2020}%
  \BibitemOpen
\bibfield  {number} {  }\bibfield  {author} {\bibinfo {author} {\bibfnamefont
  {Y.}~\bibnamefont {Gu}}, \bibinfo {author} {\bibfnamefont {A.}~\bibnamefont
  {Kitaev}}, \bibinfo {author} {\bibfnamefont {S.}~\bibnamefont {Sachdev}},\
  and\ \bibinfo {author} {\bibfnamefont {G.}~\bibnamefont {Tarnopolsky}},\
  }\bibfield  {title} {\bibinfo {title} {Notes on the complex sachdev-ye-kitaev
  model},\ }\href {http://dx.doi.org/10.1007/JHEP02(2020)157} {\bibfield
  {journal} {\bibinfo  {journal} {J. High Energy Phys.}\ }\textbf {\bibinfo
  {volume} {2020}}\bibinfo  {number} { (02)}}\BibitemShut {NoStop}%
\bibitem [{\citenamefont {Han}\ and\ \citenamefont
  {Hartnoll}(2019)}]{han2019quantum}%
  \BibitemOpen
\bibfield  {number} {  }\bibfield  {author} {\bibinfo {author} {\bibfnamefont
  {X.}~\bibnamefont {Han}}\ and\ \bibinfo {author} {\bibfnamefont {S.~A.}\
  \bibnamefont {Hartnoll}},\ }\bibfield  {title} {\bibinfo {title} {Quantum
  scrambling and state dependence of the butterfly velocity},\ }\href
  {https://scipost.org/10.21468/SciPostPhys.7.4.045} {\bibfield  {journal}
  {\bibinfo  {journal} {SciPost Phys.}\ }\textbf {\bibinfo {volume} {7}},\
  \bibinfo {pages} {045} (\bibinfo {year} {2019})}\BibitemShut {NoStop}%
\bibitem [{\citenamefont {Roberts}\ and\ \citenamefont
  {Swingle}(2016)}]{roberts2016lieb}%
  \BibitemOpen
  \bibfield  {author} {\bibinfo {author} {\bibfnamefont {D.~A.}\ \bibnamefont
  {Roberts}}\ and\ \bibinfo {author} {\bibfnamefont {B.}~\bibnamefont
  {Swingle}},\ }\bibfield  {title} {\bibinfo {title} {Lieb-robinson bound and
  the butterfly effect in quantum field theories},\ }\href
  {https://link.aps.org/doi/10.1103/PhysRevLett.117.091602} {\bibfield
  {journal} {\bibinfo  {journal} {Phys. Rev. Lett.}\ }\textbf {\bibinfo
  {volume} {117}},\ \bibinfo {pages} {091602} (\bibinfo {year}
  {2016})}\BibitemShut {NoStop}%
\bibitem [{\citenamefont {Aleiner}\ \emph {et~al.}(2016)\citenamefont
  {Aleiner}, \citenamefont {Faoro},\ and\ \citenamefont
  {Ioffe}}]{aleiner2016microscopic}%
  \BibitemOpen
  \bibfield  {author} {\bibinfo {author} {\bibfnamefont {I.~L.}\ \bibnamefont
  {Aleiner}}, \bibinfo {author} {\bibfnamefont {L.}~\bibnamefont {Faoro}},\
  and\ \bibinfo {author} {\bibfnamefont {L.~B.}\ \bibnamefont {Ioffe}},\
  }\bibfield  {title} {\bibinfo {title} {Microscopic model of quantum butterfly
  effect: out-of-time-order correlators and traveling combustion waves},\
  }\href {https://www.sciencedirect.com/journal/annals-of-physics} {\bibfield
  {journal} {\bibinfo  {journal} {Annals of Physics}\ }\textbf {\bibinfo
  {volume} {375}},\ \bibinfo {pages} {378} (\bibinfo {year}
  {2016})}\BibitemShut {NoStop}%
\bibitem [{\citenamefont {Luitz}\ and\ \citenamefont
  {Lev}(2017)}]{luitz2017information}%
  \BibitemOpen
  \bibfield  {author} {\bibinfo {author} {\bibfnamefont {D.~J.}\ \bibnamefont
  {Luitz}}\ and\ \bibinfo {author} {\bibfnamefont {Y.~B.}\ \bibnamefont
  {Lev}},\ }\bibfield  {title} {\bibinfo {title} {Information propagation in
  isolated quantum systems},\ }\href
  {https://link.aps.org/doi/10.1103/PhysRevB.96.020406} {\bibfield  {journal}
  {\bibinfo  {journal} {Phys. Rev. B}\ }\textbf {\bibinfo {volume} {96}},\
  \bibinfo {pages} {020406} (\bibinfo {year} {2017})}\BibitemShut {NoStop}%
\bibitem [{\citenamefont {Khemani}\ \emph
  {et~al.}(2018{\natexlab{a}})\citenamefont {Khemani}, \citenamefont {Huse},\
  and\ \citenamefont {Nahum}}]{khemani2018velocity}%
  \BibitemOpen
  \bibfield  {author} {\bibinfo {author} {\bibfnamefont {V.}~\bibnamefont
  {Khemani}}, \bibinfo {author} {\bibfnamefont {D.~A.}\ \bibnamefont {Huse}},\
  and\ \bibinfo {author} {\bibfnamefont {A.}~\bibnamefont {Nahum}},\ }\bibfield
   {title} {\bibinfo {title} {Velocity-dependent lyapunov exponents in
  many-body quantum, semiclassical, and classical chaos},\ }\href
  {https://link.aps.org/doi/10.1103/PhysRevB.98.144304} {\bibfield  {journal}
  {\bibinfo  {journal} {Phys. Rev. B}\ }\textbf {\bibinfo {volume} {98}},\
  \bibinfo {pages} {144304} (\bibinfo {year} {2018}{\natexlab{a}})}\BibitemShut
  {NoStop}%
\bibitem [{\citenamefont {Xu}\ and\ \citenamefont
  {Swingle}(2019{\natexlab{a}})}]{Xu_2019_nature}%
  \BibitemOpen
  \bibfield  {author} {\bibinfo {author} {\bibfnamefont {S.}~\bibnamefont
  {Xu}}\ and\ \bibinfo {author} {\bibfnamefont {B.}~\bibnamefont {Swingle}},\
  }\bibfield  {title} {\bibinfo {title} {Accessing scrambling using matrix
  product operators},\ }\href {https://doi.org/10.1038/s41567-019-0712-4}
  {\bibfield  {journal} {\bibinfo  {journal} {Nat. Phys.}\ }\textbf {\bibinfo
  {volume} {16}},\ \bibinfo {pages} {199} (\bibinfo {year}
  {2019}{\natexlab{a}})}\BibitemShut {NoStop}%
\bibitem [{\citenamefont {Lin}\ and\ \citenamefont
  {Motrunich}(2018)}]{lin2018out}%
  \BibitemOpen
  \bibfield  {author} {\bibinfo {author} {\bibfnamefont {C.-J.}\ \bibnamefont
  {Lin}}\ and\ \bibinfo {author} {\bibfnamefont {O.~I.}\ \bibnamefont
  {Motrunich}},\ }\bibfield  {title} {\bibinfo {title} {Out-of-time-ordered
  correlators in a quantum ising chain},\ }\href
  {https://link.aps.org/doi/10.1103/PhysRevB.97.144304} {\bibfield  {journal}
  {\bibinfo  {journal} {Phys. Rev. B}\ }\textbf {\bibinfo {volume} {97}},\
  \bibinfo {pages} {144304} (\bibinfo {year} {2018})}\BibitemShut {NoStop}%
\bibitem [{\citenamefont {Li}\ \emph {et~al.}(2017)\citenamefont {Li},
  \citenamefont {Fan}, \citenamefont {Wang}, \citenamefont {Ye}, \citenamefont
  {Zeng}, \citenamefont {Zhai}, \citenamefont {Peng},\ and\ \citenamefont
  {Du}}]{Li_2017}%
  \BibitemOpen
  \bibfield  {author} {\bibinfo {author} {\bibfnamefont {J.}~\bibnamefont
  {Li}}, \bibinfo {author} {\bibfnamefont {R.}~\bibnamefont {Fan}}, \bibinfo
  {author} {\bibfnamefont {H.}~\bibnamefont {Wang}}, \bibinfo {author}
  {\bibfnamefont {B.}~\bibnamefont {Ye}}, \bibinfo {author} {\bibfnamefont
  {B.}~\bibnamefont {Zeng}}, \bibinfo {author} {\bibfnamefont {H.}~\bibnamefont
  {Zhai}}, \bibinfo {author} {\bibfnamefont {X.}~\bibnamefont {Peng}},\ and\
  \bibinfo {author} {\bibfnamefont {J.}~\bibnamefont {Du}},\ }\bibfield
  {title} {\bibinfo {title} {Measuring out-of-time-order correlators on a
  nuclear magnetic resonance quantum simulator},\ }\href
  {http://dx.doi.org/10.1103/PhysRevX.7.031011} {\bibfield  {journal} {\bibinfo
   {journal} {Phys. Rev. X}\ }\textbf {\bibinfo {volume} {7}},\ \bibinfo
  {pages} {031011} (\bibinfo {year} {2017})}\BibitemShut {NoStop}%
\bibitem [{\citenamefont {Wei}\ \emph {et~al.}(2018)\citenamefont {Wei},
  \citenamefont {Ramanathan},\ and\ \citenamefont
  {Cappellaro}}]{wei2018exploring}%
  \BibitemOpen
  \bibfield  {author} {\bibinfo {author} {\bibfnamefont {K.~X.}\ \bibnamefont
  {Wei}}, \bibinfo {author} {\bibfnamefont {C.}~\bibnamefont {Ramanathan}},\
  and\ \bibinfo {author} {\bibfnamefont {P.}~\bibnamefont {Cappellaro}},\
  }\bibfield  {title} {\bibinfo {title} {Exploring localization in nuclear spin
  chains},\ }\href {https://link.aps.org/doi/10.1103/PhysRevLett.120.070501}
  {\bibfield  {journal} {\bibinfo  {journal} {Phys. Rev. Lett.}\ }\textbf
  {\bibinfo {volume} {120}},\ \bibinfo {pages} {070501} (\bibinfo {year}
  {2018})}\BibitemShut {NoStop}%
\bibitem [{\citenamefont {Nie}\ \emph {et~al.}(2019)\citenamefont {Nie},
  \citenamefont {Zhang}, \citenamefont {Zhao}, \citenamefont {Xin},
  \citenamefont {Lu},\ and\ \citenamefont {Li}}]{nie2019detecting}%
  \BibitemOpen
  \bibfield  {author} {\bibinfo {author} {\bibfnamefont {X.}~\bibnamefont
  {Nie}}, \bibinfo {author} {\bibfnamefont {Z.}~\bibnamefont {Zhang}}, \bibinfo
  {author} {\bibfnamefont {X.}~\bibnamefont {Zhao}}, \bibinfo {author}
  {\bibfnamefont {T.}~\bibnamefont {Xin}}, \bibinfo {author} {\bibfnamefont
  {D.}~\bibnamefont {Lu}},\ and\ \bibinfo {author} {\bibfnamefont
  {J.}~\bibnamefont {Li}},\ }\bibfield  {title} {\bibinfo {title} {Detecting
  scrambling via statistical correlations between randomized measurements on an
  {NMR} quantum simulator},\ }\href {https://arxiv.org/abs/1903.12237}
  {\bibfield  {journal} {\bibinfo  {journal} {arXiv:1903.12237}\ } (\bibinfo
  {year} {2019})}\BibitemShut {NoStop}%
\bibitem [{\citenamefont {S{\'a}nchez}\ \emph {et~al.}(2020)\citenamefont
  {S{\'a}nchez}, \citenamefont {Chattah}, \citenamefont {Wei}, \citenamefont
  {Buljubasich}, \citenamefont {Cappellaro},\ and\ \citenamefont
  {Pastawski}}]{sanchez2020perturbation}%
  \BibitemOpen
  \bibfield  {author} {\bibinfo {author} {\bibfnamefont {C.}~\bibnamefont
  {S{\'a}nchez}}, \bibinfo {author} {\bibfnamefont {A.}~\bibnamefont
  {Chattah}}, \bibinfo {author} {\bibfnamefont {K.}~\bibnamefont {Wei}},
  \bibinfo {author} {\bibfnamefont {L.}~\bibnamefont {Buljubasich}}, \bibinfo
  {author} {\bibfnamefont {P.}~\bibnamefont {Cappellaro}},\ and\ \bibinfo
  {author} {\bibfnamefont {H.}~\bibnamefont {Pastawski}},\ }\bibfield  {title}
  {\bibinfo {title} {Perturbation independent decay of the loschmidt echo in a
  many-body system},\ }\href
  {https://link.aps.org/doi/10.1103/PhysRevLett.124.030601} {\bibfield
  {journal} {\bibinfo  {journal} {Phys. Rev. Lett.}\ }\textbf {\bibinfo
  {volume} {124}},\ \bibinfo {pages} {030601} (\bibinfo {year}
  {2020})}\BibitemShut {NoStop}%
\bibitem [{\citenamefont {Geller}(2018)}]{Geller_2018}%
  \BibitemOpen
  \bibfield  {author} {\bibinfo {author} {\bibfnamefont {M.~R.}\ \bibnamefont
  {Geller}},\ }\bibfield  {title} {\bibinfo {title} {Sampling and scrambling on
  a chain of superconducting qubits},\ }\href
  {http://dx.doi.org/10.1103/PhysRevApplied.10.024052} {\bibfield  {journal}
  {\bibinfo  {journal} {Phys. Rev. Applied}\ }\textbf {\bibinfo {volume}
  {10}},\ \bibinfo {pages} {024052} (\bibinfo {year} {2018})}\BibitemShut
  {NoStop}%
\bibitem [{\citenamefont {Braum{\"u}ller}\ \emph {et~al.}(2022)\citenamefont
  {Braum{\"u}ller}, \citenamefont {Karamlou}, \citenamefont {Yanay},
  \citenamefont {Kannan}, \citenamefont {Kim}, \citenamefont {Kjaergaard},
  \citenamefont {Melville}, \citenamefont {Niedzielski}, \citenamefont {Sung},
  \citenamefont {Veps{\"a}l{\"a}inen} \emph {et~al.}}]{braumuller2022probing}%
  \BibitemOpen
  \bibfield  {author} {\bibinfo {author} {\bibfnamefont {J.}~\bibnamefont
  {Braum{\"u}ller}}, \bibinfo {author} {\bibfnamefont {A.~H.}\ \bibnamefont
  {Karamlou}}, \bibinfo {author} {\bibfnamefont {Y.}~\bibnamefont {Yanay}},
  \bibinfo {author} {\bibfnamefont {B.}~\bibnamefont {Kannan}}, \bibinfo
  {author} {\bibfnamefont {D.}~\bibnamefont {Kim}}, \bibinfo {author}
  {\bibfnamefont {M.}~\bibnamefont {Kjaergaard}}, \bibinfo {author}
  {\bibfnamefont {A.}~\bibnamefont {Melville}}, \bibinfo {author}
  {\bibfnamefont {B.~M.}\ \bibnamefont {Niedzielski}}, \bibinfo {author}
  {\bibfnamefont {Y.}~\bibnamefont {Sung}}, \bibinfo {author} {\bibfnamefont
  {A.}~\bibnamefont {Veps{\"a}l{\"a}inen}}, \emph {et~al.},\ }\bibfield
  {title} {\bibinfo {title} {Probing quantum information propagation with
  out-of-time-ordered correlators},\ }\href
  {https://doi.org/10.1038/s41567-021-01430-w} {\bibfield  {journal} {\bibinfo
  {journal} {Nature Physics}\ }\textbf {\bibinfo {volume} {18}},\ \bibinfo
  {pages} {172} (\bibinfo {year} {2022})}\BibitemShut {NoStop}%
\bibitem [{\citenamefont {Mi}\ \emph {et~al.}(2021)\citenamefont {Mi},
  \citenamefont {Roushan}, \citenamefont {Quintana}, \citenamefont
  {Mandr{\`a}}, \citenamefont {Marshall}, \citenamefont {Neill}, \citenamefont
  {Arute}, \citenamefont {Arya}, \citenamefont {Atalaya}, \citenamefont
  {Babbush} \emph {et~al.}}]{mi2021information}%
  \BibitemOpen
  \bibfield  {author} {\bibinfo {author} {\bibfnamefont {X.}~\bibnamefont
  {Mi}}, \bibinfo {author} {\bibfnamefont {P.}~\bibnamefont {Roushan}},
  \bibinfo {author} {\bibfnamefont {C.}~\bibnamefont {Quintana}}, \bibinfo
  {author} {\bibfnamefont {S.}~\bibnamefont {Mandr{\`a}}}, \bibinfo {author}
  {\bibfnamefont {J.}~\bibnamefont {Marshall}}, \bibinfo {author}
  {\bibfnamefont {C.}~\bibnamefont {Neill}}, \bibinfo {author} {\bibfnamefont
  {F.}~\bibnamefont {Arute}}, \bibinfo {author} {\bibfnamefont
  {K.}~\bibnamefont {Arya}}, \bibinfo {author} {\bibfnamefont {J.}~\bibnamefont
  {Atalaya}}, \bibinfo {author} {\bibfnamefont {R.}~\bibnamefont {Babbush}},
  \emph {et~al.},\ }\bibfield  {title} {\bibinfo {title} {Information
  scrambling in quantum circuits},\ }\href
  {https://www.science.org/doi/10.1126/science.abg5029} {\bibfield  {journal}
  {\bibinfo  {journal} {Science}\ }\textbf {\bibinfo {volume} {374}},\ \bibinfo
  {pages} {1479} (\bibinfo {year} {2021})}\BibitemShut {NoStop}%
\bibitem [{\citenamefont {Blok}\ \emph {et~al.}(2021)\citenamefont {Blok},
  \citenamefont {Ramasesh}, \citenamefont {Schuster}, \citenamefont
  {O’Brien}, \citenamefont {Kreikebaum}, \citenamefont {Dahlen},
  \citenamefont {Morvan}, \citenamefont {Yoshida}, \citenamefont {Yao},\ and\
  \citenamefont {Siddiqi}}]{Blok_2021}%
  \BibitemOpen
  \bibfield  {author} {\bibinfo {author} {\bibfnamefont {M.}~\bibnamefont
  {Blok}}, \bibinfo {author} {\bibfnamefont {V.}~\bibnamefont {Ramasesh}},
  \bibinfo {author} {\bibfnamefont {T.}~\bibnamefont {Schuster}}, \bibinfo
  {author} {\bibfnamefont {K.}~\bibnamefont {O’Brien}}, \bibinfo {author}
  {\bibfnamefont {J.}~\bibnamefont {Kreikebaum}}, \bibinfo {author}
  {\bibfnamefont {D.}~\bibnamefont {Dahlen}}, \bibinfo {author} {\bibfnamefont
  {A.}~\bibnamefont {Morvan}}, \bibinfo {author} {\bibfnamefont
  {B.}~\bibnamefont {Yoshida}}, \bibinfo {author} {\bibfnamefont
  {N.}~\bibnamefont {Yao}},\ and\ \bibinfo {author} {\bibfnamefont
  {I.}~\bibnamefont {Siddiqi}},\ }\bibfield  {title} {\bibinfo {title} {Quantum
  information scrambling on a superconducting qutrit processor},\ }\href
  {https://link.aps.org/doi/10.1103/PhysRevX.11.021010} {\bibfield  {journal}
  {\bibinfo  {journal} {Phys. Rev. X}\ }\textbf {\bibinfo {volume} {11}},\
  \bibinfo {pages} {021010} (\bibinfo {year} {2021})}\BibitemShut {NoStop}%
\bibitem [{\citenamefont {Gärttner}\ \emph {et~al.}(2017)\citenamefont
  {Gärttner}, \citenamefont {Bohnet}, \citenamefont {Safavi-Naini},
  \citenamefont {Wall}, \citenamefont {Bollinger},\ and\ \citenamefont
  {Rey}}]{Garttner_2017}%
  \BibitemOpen
  \bibfield  {author} {\bibinfo {author} {\bibfnamefont {M.}~\bibnamefont
  {Gärttner}}, \bibinfo {author} {\bibfnamefont {J.~G.}\ \bibnamefont
  {Bohnet}}, \bibinfo {author} {\bibfnamefont {A.}~\bibnamefont
  {Safavi-Naini}}, \bibinfo {author} {\bibfnamefont {M.~L.}\ \bibnamefont
  {Wall}}, \bibinfo {author} {\bibfnamefont {J.~J.}\ \bibnamefont
  {Bollinger}},\ and\ \bibinfo {author} {\bibfnamefont {A.~M.}\ \bibnamefont
  {Rey}},\ }\bibfield  {title} {\bibinfo {title} {Measuring out-of-time-order
  correlations and multiple quantum spectra in a trapped-ion quantum magnet},\
  }\href {http://dx.doi.org/10.1038/nphys4119} {\bibfield  {journal} {\bibinfo
  {journal} {Nat. Phys.}\ }\textbf {\bibinfo {volume} {13}},\ \bibinfo {pages}
  {781} (\bibinfo {year} {2017})}\BibitemShut {NoStop}%
\bibitem [{\citenamefont {Landsman}\ \emph {et~al.}(2019)\citenamefont
  {Landsman}, \citenamefont {Figgatt}, \citenamefont {Schuster}, \citenamefont
  {Linke}, \citenamefont {Yoshida}, \citenamefont {Yao},\ and\ \citenamefont
  {Monroe}}]{Landsman_2019}%
  \BibitemOpen
  \bibfield  {author} {\bibinfo {author} {\bibfnamefont {K.~A.}\ \bibnamefont
  {Landsman}}, \bibinfo {author} {\bibfnamefont {C.}~\bibnamefont {Figgatt}},
  \bibinfo {author} {\bibfnamefont {T.}~\bibnamefont {Schuster}}, \bibinfo
  {author} {\bibfnamefont {N.~M.}\ \bibnamefont {Linke}}, \bibinfo {author}
  {\bibfnamefont {B.}~\bibnamefont {Yoshida}}, \bibinfo {author} {\bibfnamefont
  {N.~Y.}\ \bibnamefont {Yao}},\ and\ \bibinfo {author} {\bibfnamefont
  {C.}~\bibnamefont {Monroe}},\ }\bibfield  {title} {\bibinfo {title} {Verified
  quantum information scrambling},\ }\href
  {http://dx.doi.org/10.1038/s41586-019-0952-6} {\bibfield  {journal} {\bibinfo
   {journal} {Nature}\ }\textbf {\bibinfo {volume} {567}},\ \bibinfo {pages}
  {61} (\bibinfo {year} {2019})}\BibitemShut {NoStop}%
\bibitem [{\citenamefont {Joshi}\ \emph {et~al.}(2020)\citenamefont {Joshi},
  \citenamefont {Elben}, \citenamefont {Vermersch}, \citenamefont {Brydges},
  \citenamefont {Maier}, \citenamefont {Zoller}, \citenamefont {Blatt},\ and\
  \citenamefont {Roos}}]{joshi2020quantum}%
  \BibitemOpen
  \bibfield  {author} {\bibinfo {author} {\bibfnamefont {M.~K.}\ \bibnamefont
  {Joshi}}, \bibinfo {author} {\bibfnamefont {A.}~\bibnamefont {Elben}},
  \bibinfo {author} {\bibfnamefont {B.}~\bibnamefont {Vermersch}}, \bibinfo
  {author} {\bibfnamefont {T.}~\bibnamefont {Brydges}}, \bibinfo {author}
  {\bibfnamefont {C.}~\bibnamefont {Maier}}, \bibinfo {author} {\bibfnamefont
  {P.}~\bibnamefont {Zoller}}, \bibinfo {author} {\bibfnamefont
  {R.}~\bibnamefont {Blatt}},\ and\ \bibinfo {author} {\bibfnamefont {C.~F.}\
  \bibnamefont {Roos}},\ }\bibfield  {title} {\bibinfo {title} {Quantum
  information scrambling in a trapped-ion quantum simulator with tunable range
  interactions},\ }\href
  {https://link.aps.org/doi/10.1103/PhysRevLett.124.240505} {\bibfield
  {journal} {\bibinfo  {journal} {Phys. Rev. Lett.}\ }\textbf {\bibinfo
  {volume} {124}},\ \bibinfo {pages} {240505} (\bibinfo {year}
  {2020})}\BibitemShut {NoStop}%
\bibitem [{\citenamefont {Xu}\ and\ \citenamefont
  {Swingle}(2019{\natexlab{b}})}]{Xu_2019}%
  \BibitemOpen
  \bibfield  {author} {\bibinfo {author} {\bibfnamefont {S.}~\bibnamefont
  {Xu}}\ and\ \bibinfo {author} {\bibfnamefont {B.}~\bibnamefont {Swingle}},\
  }\bibfield  {title} {\bibinfo {title} {Locality, quantum fluctuations, and
  scrambling},\ }\href {http://doi.org/10.1103/PhysRevX.9.031048} {\bibfield
  {journal} {\bibinfo  {journal} {Phys. Rev. X}\ }\textbf {\bibinfo {volume}
  {9}},\ \bibinfo {pages} {031048} (\bibinfo {year}
  {2019}{\natexlab{b}})}\BibitemShut {NoStop}%
\bibitem [{\citenamefont {Qi}\ and\ \citenamefont
  {Streicher}(2019)}]{qi2019quantum}%
  \BibitemOpen
  \bibfield  {author} {\bibinfo {author} {\bibfnamefont {X.-L.}\ \bibnamefont
  {Qi}}\ and\ \bibinfo {author} {\bibfnamefont {A.}~\bibnamefont {Streicher}},\
  }\bibfield  {title} {\bibinfo {title} {Quantum epidemiology: operator growth,
  thermal effects, and syk},\ }\href {https://doi.org/10.1007/JHEP08(2019)012}
  {\bibfield  {journal} {\bibinfo  {journal} {J. High Energy Phys.}\ }\textbf
  {\bibinfo {volume} {2019}}\bibinfo  {number} { (8)},\ \bibinfo {pages}
  {1}}\BibitemShut {NoStop}%
\bibitem [{\citenamefont {Zhou}\ and\ \citenamefont
  {Swingle}(2021)}]{zhou2021operator}%
  \BibitemOpen
\bibfield  {number} {  }\bibfield  {author} {\bibinfo {author} {\bibfnamefont
  {T.}~\bibnamefont {Zhou}}\ and\ \bibinfo {author} {\bibfnamefont
  {B.}~\bibnamefont {Swingle}},\ }\bibfield  {title} {\bibinfo {title}
  {Operator growth from global out-of-time-order correlators},\ }\href
  {https://arxiv.org/abs/2112.01562} {\bibfield  {journal} {\bibinfo  {journal}
  {arXiv:2112.01562}\ } (\bibinfo {year} {2021})}\BibitemShut {NoStop}%
\bibitem [{\citenamefont {Zhou}\ \emph {et~al.}(2022)\citenamefont {Zhou},
  \citenamefont {Su}, \citenamefont {Halimeh}, \citenamefont {Ott},
  \citenamefont {Sun}, \citenamefont {Hauke}, \citenamefont {Yang},
  \citenamefont {Yuan}, \citenamefont {Berges},\ and\ \citenamefont
  {Pan}}]{Zhou_charged_2022}%
  \BibitemOpen
  \bibfield  {author} {\bibinfo {author} {\bibfnamefont {Z.-Y.}\ \bibnamefont
  {Zhou}}, \bibinfo {author} {\bibfnamefont {G.-X.}\ \bibnamefont {Su}},
  \bibinfo {author} {\bibfnamefont {J.~C.}\ \bibnamefont {Halimeh}}, \bibinfo
  {author} {\bibfnamefont {R.}~\bibnamefont {Ott}}, \bibinfo {author}
  {\bibfnamefont {H.}~\bibnamefont {Sun}}, \bibinfo {author} {\bibfnamefont
  {P.}~\bibnamefont {Hauke}}, \bibinfo {author} {\bibfnamefont
  {B.}~\bibnamefont {Yang}}, \bibinfo {author} {\bibfnamefont {Z.-S.}\
  \bibnamefont {Yuan}}, \bibinfo {author} {\bibfnamefont {J.}~\bibnamefont
  {Berges}},\ and\ \bibinfo {author} {\bibfnamefont {J.-W.}\ \bibnamefont
  {Pan}},\ }\bibfield  {title} {\bibinfo {title} {Thermalization dynamics of a
  gauge theory on a quantum simulator},\ }\href
  {https://doi.org/10.1126/science.abl6277} {\bibfield  {journal} {\bibinfo
  {journal} {Science}\ }\textbf {\bibinfo {volume} {377}},\ \bibinfo {pages}
  {311} (\bibinfo {year} {2022})}\BibitemShut {NoStop}%
\bibitem [{\citenamefont {Patel}\ \emph {et~al.}(2017)\citenamefont {Patel},
  \citenamefont {Chowdhury}, \citenamefont {Sachdev},\ and\ \citenamefont
  {Swingle}}]{Patel_2017}%
  \BibitemOpen
  \bibfield  {author} {\bibinfo {author} {\bibfnamefont {A.~A.}\ \bibnamefont
  {Patel}}, \bibinfo {author} {\bibfnamefont {D.}~\bibnamefont {Chowdhury}},
  \bibinfo {author} {\bibfnamefont {S.}~\bibnamefont {Sachdev}},\ and\ \bibinfo
  {author} {\bibfnamefont {B.}~\bibnamefont {Swingle}},\ }\bibfield  {title}
  {\bibinfo {title} {Quantum butterfly effect in weakly interacting diffusive
  metals},\ }\href {http://doi.org/10.1103/PhysRevX.7.031047} {\bibfield
  {journal} {\bibinfo  {journal} {Phys. Rev. X}\ }\textbf {\bibinfo {volume}
  {7}},\ \bibinfo {pages} {031047} (\bibinfo {year} {2017})}\BibitemShut
  {NoStop}%
\bibitem [{\citenamefont {Khemani}\ \emph
  {et~al.}(2018{\natexlab{b}})\citenamefont {Khemani}, \citenamefont
  {Vishwanath},\ and\ \citenamefont {Huse}}]{Khemani_2018}%
  \BibitemOpen
  \bibfield  {author} {\bibinfo {author} {\bibfnamefont {V.}~\bibnamefont
  {Khemani}}, \bibinfo {author} {\bibfnamefont {A.}~\bibnamefont
  {Vishwanath}},\ and\ \bibinfo {author} {\bibfnamefont {D.~A.}\ \bibnamefont
  {Huse}},\ }\bibfield  {title} {\bibinfo {title} {Operator spreading and the
  emergence of dissipative hydrodynamics under unitary evolution with
  conservation laws},\ }\href {http://dx.doi.org/10.1103/PhysRevX.8.031057}
  {\bibfield  {journal} {\bibinfo  {journal} {Phys. Rev. X}\ }\textbf {\bibinfo
  {volume} {8}},\ \bibinfo {pages} {031057} (\bibinfo {year}
  {2018}{\natexlab{b}})}\BibitemShut {NoStop}%
\bibitem [{\citenamefont {Rakovszky}\ \emph {et~al.}(2018)\citenamefont
  {Rakovszky}, \citenamefont {Pollmann},\ and\ \citenamefont {von
  Keyserlingk}}]{Rakovszky_2018}%
  \BibitemOpen
  \bibfield  {author} {\bibinfo {author} {\bibfnamefont {T.}~\bibnamefont
  {Rakovszky}}, \bibinfo {author} {\bibfnamefont {F.}~\bibnamefont
  {Pollmann}},\ and\ \bibinfo {author} {\bibfnamefont {C.}~\bibnamefont {von
  Keyserlingk}},\ }\bibfield  {title} {\bibinfo {title} {Diffusive
  hydrodynamics of out-of-time-ordered correlators with charge conservation},\
  }\href {http://dx.doi.org/10.1103/PhysRevX.8.031058} {\bibfield  {journal}
  {\bibinfo  {journal} {Phys. Rev. X}\ }\textbf {\bibinfo {volume} {8}},\
  \bibinfo {pages} {031058} (\bibinfo {year} {2018})}\BibitemShut {NoStop}%
\bibitem [{\citenamefont {Hunter-Jones}(2018)}]{hunter2018operator}%
  \BibitemOpen
  \bibfield  {author} {\bibinfo {author} {\bibfnamefont {N.}~\bibnamefont
  {Hunter-Jones}},\ }\bibfield  {title} {\bibinfo {title} {Operator growth in
  random quantum circuits with symmetry},\ }\href
  {https://arxiv.org/abs/1812.08219} {\bibfield  {journal} {\bibinfo  {journal}
  {arXiv:1812.08219}\ } (\bibinfo {year} {2018})}\BibitemShut {NoStop}%
\bibitem [{\citenamefont {Chen}\ \emph
  {et~al.}(2020{\natexlab{a}})\citenamefont {Chen}, \citenamefont
  {Nandkishore},\ and\ \citenamefont {Lucas}}]{chen2020quantum}%
  \BibitemOpen
  \bibfield  {author} {\bibinfo {author} {\bibfnamefont {X.}~\bibnamefont
  {Chen}}, \bibinfo {author} {\bibfnamefont {R.~M.}\ \bibnamefont
  {Nandkishore}},\ and\ \bibinfo {author} {\bibfnamefont {A.}~\bibnamefont
  {Lucas}},\ }\bibfield  {title} {\bibinfo {title} {Quantum butterfly effect in
  polarized floquet systems},\ }\href
  {https://link.aps.org/doi/10.1103/PhysRevB.101.064307} {\bibfield  {journal}
  {\bibinfo  {journal} {Phys. Rev. B}\ }\textbf {\bibinfo {volume} {101}},\
  \bibinfo {pages} {064307} (\bibinfo {year} {2020}{\natexlab{a}})}\BibitemShut
  {NoStop}%
\bibitem [{\citenamefont {Chen}\ \emph
  {et~al.}(2020{\natexlab{b}})\citenamefont {Chen}, \citenamefont {Gu},\ and\
  \citenamefont {Lucas}}]{Chen_2020}%
  \BibitemOpen
  \bibfield  {author} {\bibinfo {author} {\bibfnamefont {X.}~\bibnamefont
  {Chen}}, \bibinfo {author} {\bibfnamefont {Y.}~\bibnamefont {Gu}},\ and\
  \bibinfo {author} {\bibfnamefont {A.}~\bibnamefont {Lucas}},\ }\bibfield
  {title} {\bibinfo {title} {Many-body quantum dynamics slows down at low
  density},\ }\href {https://scipost.org/10.21468/SciPostPhys.9.5.071}
  {\bibfield  {journal} {\bibinfo  {journal} {SciPost Phys.}\ }\textbf
  {\bibinfo {volume} {9}},\ \bibinfo {pages} {71} (\bibinfo {year}
  {2020}{\natexlab{b}})}\BibitemShut {NoStop}%
\bibitem [{\citenamefont {Bao}\ \emph {et~al.}(2021)\citenamefont {Bao},
  \citenamefont {Choi},\ and\ \citenamefont {Altman}}]{bao2021symmetry}%
  \BibitemOpen
  \bibfield  {author} {\bibinfo {author} {\bibfnamefont {Y.}~\bibnamefont
  {Bao}}, \bibinfo {author} {\bibfnamefont {S.}~\bibnamefont {Choi}},\ and\
  \bibinfo {author} {\bibfnamefont {E.}~\bibnamefont {Altman}},\ }\bibfield
  {title} {\bibinfo {title} {Symmetry enriched phases of quantum circuits},\
  }\href {https://doi.org/10.1016/j.aop.2021.168618} {\bibfield  {journal}
  {\bibinfo  {journal} {Annals of Physics}\ }\textbf {\bibinfo {volume}
  {435}},\ \bibinfo {pages} {168618} (\bibinfo {year} {2021})}\BibitemShut
  {NoStop}%
\bibitem [{\citenamefont {Agarwal}\ and\ \citenamefont
  {Xu}(2022)}]{agarwal2022emergent}%
  \BibitemOpen
  \bibfield  {author} {\bibinfo {author} {\bibfnamefont {L.}~\bibnamefont
  {Agarwal}}\ and\ \bibinfo {author} {\bibfnamefont {S.}~\bibnamefont {Xu}},\
  }\bibfield  {title} {\bibinfo {title} {Emergent symmetry in brownian syk
  models and charge dependent scrambling},\ }\href
  {https://doi.org/10.1007/JHEP02(2022)045} {\bibfield  {journal} {\bibinfo
  {journal} {J. High Energy Phys.}\ }\textbf {\bibinfo {volume} {2022}}\bibinfo
   {number} { (2)},\ \bibinfo {pages} {1}}\BibitemShut {NoStop}%
\bibitem [{\citenamefont {Gu}\ \emph {et~al.}(2017)\citenamefont {Gu},
  \citenamefont {Lucas},\ and\ \citenamefont {Qi}}]{gu2017energy}%
  \BibitemOpen
\bibfield  {number} {  }\bibfield  {author} {\bibinfo {author} {\bibfnamefont
  {Y.}~\bibnamefont {Gu}}, \bibinfo {author} {\bibfnamefont {A.}~\bibnamefont
  {Lucas}},\ and\ \bibinfo {author} {\bibfnamefont {X.-L.}\ \bibnamefont
  {Qi}},\ }\bibfield  {title} {\bibinfo {title} {Energy diffusion and the
  butterfly effect in inhomogeneous sachdev-ye-kitaev chains},\ }\href
  {https://scipost.org/10.21468/SciPostPhys.2.3.018} {\bibfield  {journal}
  {\bibinfo  {journal} {SciPost Physics}\ }\textbf {\bibinfo {volume} {2}},\
  \bibinfo {pages} {018} (\bibinfo {year} {2017})}\BibitemShut {NoStop}%
\bibitem [{\citenamefont {Pai}\ \emph {et~al.}(2019)\citenamefont {Pai},
  \citenamefont {Pretko},\ and\ \citenamefont
  {Nandkishore}}]{pai2019localization}%
  \BibitemOpen
  \bibfield  {author} {\bibinfo {author} {\bibfnamefont {S.}~\bibnamefont
  {Pai}}, \bibinfo {author} {\bibfnamefont {M.}~\bibnamefont {Pretko}},\ and\
  \bibinfo {author} {\bibfnamefont {R.~M.}\ \bibnamefont {Nandkishore}},\
  }\bibfield  {title} {\bibinfo {title} {Localization in fractonic random
  circuits},\ }\href {https://link.aps.org/doi/10.1103/PhysRevX.9.021003}
  {\bibfield  {journal} {\bibinfo  {journal} {Phys. Rev. X}\ }\textbf {\bibinfo
  {volume} {9}},\ \bibinfo {pages} {021003} (\bibinfo {year}
  {2019})}\BibitemShut {NoStop}%
\bibitem [{\citenamefont {Feldmeier}\ and\ \citenamefont
  {Knap}(2021)}]{feldmeier2021critically}%
  \BibitemOpen
  \bibfield  {author} {\bibinfo {author} {\bibfnamefont {J.}~\bibnamefont
  {Feldmeier}}\ and\ \bibinfo {author} {\bibfnamefont {M.}~\bibnamefont
  {Knap}},\ }\bibfield  {title} {\bibinfo {title} {Critically slow operator
  dynamics in constrained many-body systems},\ }\href
  {https://link.aps.org/doi/10.1103/PhysRevLett.127.235301} {\bibfield
  {journal} {\bibinfo  {journal} {Phys. Rev. Lett.}\ }\textbf {\bibinfo
  {volume} {127}},\ \bibinfo {pages} {235301} (\bibinfo {year}
  {2021})}\BibitemShut {NoStop}%
\bibitem [{\citenamefont {Cotler}\ and\ \citenamefont
  {Hunter-Jones}(2020)}]{cotler2020spectral}%
  \BibitemOpen
  \bibfield  {author} {\bibinfo {author} {\bibfnamefont {J.}~\bibnamefont
  {Cotler}}\ and\ \bibinfo {author} {\bibfnamefont {N.}~\bibnamefont
  {Hunter-Jones}},\ }\bibfield  {title} {\bibinfo {title} {Spectral decoupling
  in many-body quantum chaos},\ }\href
  {https://doi.org/10.1007/JHEP12(2020)205} {\bibfield  {journal} {\bibinfo
  {journal} {J. High Energy Phys.}\ }\textbf {\bibinfo {volume} {2020}}\bibinfo
   {number} { (12)},\ \bibinfo {pages} {1}}\BibitemShut {NoStop}%
\bibitem [{\citenamefont {Zanoci}\ and\ \citenamefont
  {Swingle}(2022)}]{zanoci2022near}%
  \BibitemOpen
\bibfield  {number} {  }\bibfield  {author} {\bibinfo {author} {\bibfnamefont
  {C.}~\bibnamefont {Zanoci}}\ and\ \bibinfo {author} {\bibfnamefont
  {B.}~\bibnamefont {Swingle}},\ }\bibfield  {title} {\bibinfo {title}
  {Near-equilibrium approach to transport in complex sachdev-ye-kitaev
  models},\ }\href {https://doi.org/10.1103/PhysRevB.105.235131} {\bibfield
  {journal} {\bibinfo  {journal} {Phys. Rev. B}\ }\textbf {\bibinfo {volume}
  {105}},\ \bibinfo {pages} {235131} (\bibinfo {year} {2022})}\BibitemShut
  {NoStop}%
\bibitem [{\citenamefont {Friedman}\ \emph {et~al.}(2019)\citenamefont
  {Friedman}, \citenamefont {Chan}, \citenamefont {De~Luca},\ and\
  \citenamefont {Chalker}}]{friedman2019spectral}%
  \BibitemOpen
  \bibfield  {author} {\bibinfo {author} {\bibfnamefont {A.~J.}\ \bibnamefont
  {Friedman}}, \bibinfo {author} {\bibfnamefont {A.}~\bibnamefont {Chan}},
  \bibinfo {author} {\bibfnamefont {A.}~\bibnamefont {De~Luca}},\ and\ \bibinfo
  {author} {\bibfnamefont {J.}~\bibnamefont {Chalker}},\ }\bibfield  {title}
  {\bibinfo {title} {Spectral statistics and many-body quantum chaos with
  conserved charge},\ }\href
  {https://link.aps.org/doi/10.1103/PhysRevLett.123.210603} {\bibfield
  {journal} {\bibinfo  {journal} {Phys. Rev. Lett.}\ }\textbf {\bibinfo
  {volume} {123}},\ \bibinfo {pages} {210603} (\bibinfo {year}
  {2019})}\BibitemShut {NoStop}%
\bibitem [{\citenamefont {Roy}\ and\ \citenamefont
  {Prosen}(2020)}]{Roy2020SFF}%
  \BibitemOpen
  \bibfield  {author} {\bibinfo {author} {\bibfnamefont {D.}~\bibnamefont
  {Roy}}\ and\ \bibinfo {author} {\bibfnamefont {T.}~\bibnamefont {Prosen}},\
  }\bibfield  {title} {\bibinfo {title} {Random matrix spectral form factor in
  kicked interacting fermionic chains},\ }\href
  {https://doi.org/10.1103/PhysRevE.102.060202} {\bibfield  {journal} {\bibinfo
   {journal} {Phys. Rev. E}\ }\textbf {\bibinfo {volume} {102}},\ \bibinfo
  {pages} {060202} (\bibinfo {year} {2020})}\BibitemShut {NoStop}%
\bibitem [{\citenamefont {Kos}\ \emph {et~al.}(2021)\citenamefont {Kos},
  \citenamefont {Bertini},\ and\ \citenamefont {Prosen}}]{kos2021chaos}%
  \BibitemOpen
  \bibfield  {author} {\bibinfo {author} {\bibfnamefont {P.}~\bibnamefont
  {Kos}}, \bibinfo {author} {\bibfnamefont {B.}~\bibnamefont {Bertini}},\ and\
  \bibinfo {author} {\bibfnamefont {T.}~\bibnamefont {Prosen}},\ }\bibfield
  {title} {\bibinfo {title} {Chaos and ergodicity in extended quantum systems
  with noisy driving},\ }\href
  {https://link.aps.org/doi/10.1103/PhysRevLett.126.190601} {\bibfield
  {journal} {\bibinfo  {journal} {Phys. Rev. Lett.}\ }\textbf {\bibinfo
  {volume} {126}},\ \bibinfo {pages} {190601} (\bibinfo {year}
  {2021})}\BibitemShut {NoStop}%
\bibitem [{\citenamefont {Moudgalya}\ \emph {et~al.}(2021)\citenamefont
  {Moudgalya}, \citenamefont {Prem}, \citenamefont {Huse},\ and\ \citenamefont
  {Chan}}]{Moudgalya_2021}%
  \BibitemOpen
  \bibfield  {author} {\bibinfo {author} {\bibfnamefont {S.}~\bibnamefont
  {Moudgalya}}, \bibinfo {author} {\bibfnamefont {A.}~\bibnamefont {Prem}},
  \bibinfo {author} {\bibfnamefont {D.~A.}\ \bibnamefont {Huse}},\ and\
  \bibinfo {author} {\bibfnamefont {A.}~\bibnamefont {Chan}},\ }\bibfield
  {title} {\bibinfo {title} {Spectral statistics in constrained many-body
  quantum chaotic systems},\ }\href
  {http://dx.doi.org/10.1103/PhysRevResearch.3.023176} {\bibfield  {journal}
  {\bibinfo  {journal} {Phys. Rev. Research}\ }\textbf {\bibinfo {volume}
  {03}},\ \bibinfo {pages} {023176} (\bibinfo {year} {2021})}\BibitemShut
  {NoStop}%
\bibitem [{\citenamefont {Singh}\ \emph {et~al.}(2021)\citenamefont {Singh},
  \citenamefont {Ware}, \citenamefont {Vasseur},\ and\ \citenamefont
  {Friedman}}]{singh2021subdiffusion}%
  \BibitemOpen
  \bibfield  {author} {\bibinfo {author} {\bibfnamefont {H.}~\bibnamefont
  {Singh}}, \bibinfo {author} {\bibfnamefont {B.~A.}\ \bibnamefont {Ware}},
  \bibinfo {author} {\bibfnamefont {R.}~\bibnamefont {Vasseur}},\ and\ \bibinfo
  {author} {\bibfnamefont {A.~J.}\ \bibnamefont {Friedman}},\ }\bibfield
  {title} {\bibinfo {title} {Subdiffusion and many-body quantum chaos with
  kinetic constraints},\ }\href
  {https://link.aps.org/doi/10.1103/PhysRevLett.127.230602} {\bibfield
  {journal} {\bibinfo  {journal} {Phys. Rev. Lett.}\ }\textbf {\bibinfo
  {volume} {127}},\ \bibinfo {pages} {230602} (\bibinfo {year}
  {2021})}\BibitemShut {NoStop}%
\bibitem [{\citenamefont {Roy}\ \emph {et~al.}(2022)\citenamefont {Roy},
  \citenamefont {Mishra},\ and\ \citenamefont {Prosen}}]{roy2022spectral}%
  \BibitemOpen
  \bibfield  {author} {\bibinfo {author} {\bibfnamefont {D.}~\bibnamefont
  {Roy}}, \bibinfo {author} {\bibfnamefont {D.}~\bibnamefont {Mishra}},\ and\
  \bibinfo {author} {\bibfnamefont {T.}~\bibnamefont {Prosen}},\ }\bibfield
  {title} {\bibinfo {title} {Spectral form factor in a minimal bosonic model of
  many-body quantum chaos},\ }\href
  {https://doi.org/10.1103/PhysRevE.106.024208} {\bibfield  {journal} {\bibinfo
   {journal} {Phys. Rev. E}\ }\textbf {\bibinfo {volume} {106}},\ \bibinfo
  {pages} {024208} (\bibinfo {year} {2022})}\BibitemShut {NoStop}%
\bibitem [{\citenamefont {Golding}\ \emph {et~al.}(1998)\citenamefont
  {Golding}, \citenamefont {Kozlovsky}, \citenamefont {Cohen},\ and\
  \citenamefont {Ben-Jacob}}]{golding1998studies}%
  \BibitemOpen
  \bibfield  {author} {\bibinfo {author} {\bibfnamefont {I.}~\bibnamefont
  {Golding}}, \bibinfo {author} {\bibfnamefont {Y.}~\bibnamefont {Kozlovsky}},
  \bibinfo {author} {\bibfnamefont {I.}~\bibnamefont {Cohen}},\ and\ \bibinfo
  {author} {\bibfnamefont {E.}~\bibnamefont {Ben-Jacob}},\ }\bibfield  {title}
  {\bibinfo {title} {Studies of bacterial branching growth using
  reaction--diffusion models for colonial development},\ }\href
  {https://www.sciencedirect.com/science/article/pii/S0378437198003458}
  {\bibfield  {journal} {\bibinfo  {journal} {Physica A: Statistical Mechanics
  and its Applications}\ }\textbf {\bibinfo {volume} {260}},\ \bibinfo {pages}
  {510} (\bibinfo {year} {1998})}\BibitemShut {NoStop}%
\bibitem [{\citenamefont {Sachdev}\ and\ \citenamefont
  {Ye}(1993)}]{Sachdev1992}%
  \BibitemOpen
  \bibfield  {author} {\bibinfo {author} {\bibfnamefont {S.}~\bibnamefont
  {Sachdev}}\ and\ \bibinfo {author} {\bibfnamefont {J.}~\bibnamefont {Ye}},\
  }\bibfield  {title} {\bibinfo {title} {{Gapless spin-fluid ground state in a
  random quantum Heisenberg magnet}},\ }\href
  {http://arxiv.org/abs/cond-mat/9212030
  http://dx.doi.org/10.1103/PhysRevLett.70.3339} {\bibfield  {journal}
  {\bibinfo  {journal} {Phys. Rev. Lett.}\ }\textbf {\bibinfo {volume} {70}},\
  \bibinfo {pages} {3339} (\bibinfo {year} {1993})}\BibitemShut {NoStop}%
\bibitem [{\citenamefont {Sachdev}(2015)}]{Sachdev_2015}%
  \BibitemOpen
  \bibfield  {author} {\bibinfo {author} {\bibfnamefont {S.}~\bibnamefont
  {Sachdev}},\ }\bibfield  {title} {\bibinfo {title} {Bekenstein-hawking
  entropy and strange metals},\ }\href
  {https://link.aps.org/doi/10.1103/PhysRevX.5.041025} {\bibfield  {journal}
  {\bibinfo  {journal} {Phys. Rev. X}\ }\textbf {\bibinfo {volume} {5}},\
  \bibinfo {pages} {041025} (\bibinfo {year} {2015})}\BibitemShut {NoStop}%
\bibitem [{\citenamefont {Kitaev}(2015)}]{kitaev2015}%
  \BibitemOpen
  \bibfield  {author} {\bibinfo {author} {\bibfnamefont {A.}~\bibnamefont
  {Kitaev}},\ }\bibfield  {title} {\bibinfo {title} {{A simple model of quantum
  holography}},\ }in\ \href@noop {} {\emph {\bibinfo {booktitle} {KITP Program:
  Entanglement in Strongly-Correlated Quantum Matter}}}\ (\bibinfo {year}
  {2015})\BibitemShut {NoStop}%
\bibitem [{\citenamefont {Saad}\ \emph {et~al.}(2018)\citenamefont {Saad},
  \citenamefont {Shenker},\ and\ \citenamefont
  {Stanford}}]{saad2018semiclassical}%
  \BibitemOpen
  \bibfield  {author} {\bibinfo {author} {\bibfnamefont {P.}~\bibnamefont
  {Saad}}, \bibinfo {author} {\bibfnamefont {S.~H.}\ \bibnamefont {Shenker}},\
  and\ \bibinfo {author} {\bibfnamefont {D.}~\bibnamefont {Stanford}},\
  }\bibfield  {title} {\bibinfo {title} {A semiclassical ramp in {SYK} and in
  gravity},\ }\href {https://arxiv.org/abs/1806.06840} {\bibfield  {journal}
  {\bibinfo  {journal} {arXiv:1806.06840}\ } (\bibinfo {year}
  {2018})}\BibitemShut {NoStop}%
\bibitem [{\citenamefont {Sünderhauf}\ \emph {et~al.}(2019)\citenamefont
  {Sünderhauf}, \citenamefont {Piroli}, \citenamefont {Qi}, \citenamefont
  {Schuch},\ and\ \citenamefont {Cirac}}]{Sunderhauf2019}%
  \BibitemOpen
  \bibfield  {author} {\bibinfo {author} {\bibfnamefont {C.}~\bibnamefont
  {Sünderhauf}}, \bibinfo {author} {\bibfnamefont {L.}~\bibnamefont {Piroli}},
  \bibinfo {author} {\bibfnamefont {X.-L.}\ \bibnamefont {Qi}}, \bibinfo
  {author} {\bibfnamefont {N.}~\bibnamefont {Schuch}},\ and\ \bibinfo {author}
  {\bibfnamefont {J.~I.}\ \bibnamefont {Cirac}},\ }\bibfield  {title} {\bibinfo
  {title} {Quantum chaos in the {Brownian} {SYK} model with large finite {$N$}:
  {OTOC}s and tripartite information},\ }\href
  {http://dx.doi.org/10.1007/JHEP11(2019)038} {\bibfield  {journal} {\bibinfo
  {journal} {J. High Energy Phys.}\ }\textbf {\bibinfo {volume} {2019}}\bibinfo
   {number} { (11)},\ \bibinfo {pages} {038}}\BibitemShut {NoStop}%
\bibitem [{\citenamefont {Jian}\ \emph {et~al.}(2021)\citenamefont {Jian},
  \citenamefont {Liu}, \citenamefont {Chen}, \citenamefont {Swingle},\ and\
  \citenamefont {Zhang}}]{jian2021measurement}%
  \BibitemOpen
\bibfield  {number} {  }\bibfield  {author} {\bibinfo {author} {\bibfnamefont
  {S.-K.}\ \bibnamefont {Jian}}, \bibinfo {author} {\bibfnamefont
  {C.}~\bibnamefont {Liu}}, \bibinfo {author} {\bibfnamefont {X.}~\bibnamefont
  {Chen}}, \bibinfo {author} {\bibfnamefont {B.}~\bibnamefont {Swingle}},\ and\
  \bibinfo {author} {\bibfnamefont {P.}~\bibnamefont {Zhang}},\ }\bibfield
  {title} {\bibinfo {title} {Measurement-induced phase transition in the
  monitored sachdev-ye-kitaev model},\ }\href
  {https://link.aps.org/doi/10.1103/PhysRevLett.127.140601} {\bibfield
  {journal} {\bibinfo  {journal} {Phys. Rev. Lett.}\ }\textbf {\bibinfo
  {volume} {127}},\ \bibinfo {pages} {140601} (\bibinfo {year}
  {2021})}\BibitemShut {NoStop}%
\bibitem [{\citenamefont {Jian}\ and\ \citenamefont
  {Swingle}(2021)}]{Jian_2021}%
  \BibitemOpen
  \bibfield  {author} {\bibinfo {author} {\bibfnamefont {S.-K.}\ \bibnamefont
  {Jian}}\ and\ \bibinfo {author} {\bibfnamefont {B.}~\bibnamefont {Swingle}},\
  }\bibfield  {title} {\bibinfo {title} {Note on entropy dynamics in the
  brownian {SYK} model},\ }\href {http://dx.doi.org/10.1007/JHEP03(2021)042}
  {\bibfield  {journal} {\bibinfo  {journal} {J. High Energy Phys.}\ }\textbf
  {\bibinfo {volume} {2021}}\bibinfo  {number} { (03)},\ \bibinfo {pages}
  {042}}\BibitemShut {NoStop}%
\bibitem [{\citenamefont {Stanford}\ \emph {et~al.}(2022)\citenamefont
  {Stanford}, \citenamefont {Yang},\ and\ \citenamefont
  {Yao}}]{stanford2022subleading}%
  \BibitemOpen
\bibfield  {number} {  }\bibfield  {author} {\bibinfo {author} {\bibfnamefont
  {D.}~\bibnamefont {Stanford}}, \bibinfo {author} {\bibfnamefont
  {Z.}~\bibnamefont {Yang}},\ and\ \bibinfo {author} {\bibfnamefont
  {S.}~\bibnamefont {Yao}},\ }\bibfield  {title} {\bibinfo {title} {Subleading
  weingartens},\ }\href {https://doi.org/10.1007/JHEP02(2022)200} {\bibfield
  {journal} {\bibinfo  {journal} {J. High Energy Phys.}\ }\textbf {\bibinfo
  {volume} {2022}}\bibinfo  {number} { (2)},\ \bibinfo {pages} {1}}\BibitemShut
  {NoStop}%
\bibitem [{\citenamefont {Zhou}\ and\ \citenamefont {Chen}(2019)}]{Zhou_2019}%
  \BibitemOpen
\bibfield  {number} {  }\bibfield  {author} {\bibinfo {author} {\bibfnamefont
  {T.}~\bibnamefont {Zhou}}\ and\ \bibinfo {author} {\bibfnamefont
  {X.}~\bibnamefont {Chen}},\ }\bibfield  {title} {\bibinfo {title} {Operator
  dynamics in a brownian quantum circuit},\ }\href
  {http://dx.doi.org/10.1103/PhysRevE.99.052212} {\bibfield  {journal}
  {\bibinfo  {journal} {Phys. Rev. E}\ }\textbf {\bibinfo {volume} {99}},\
  \bibinfo {pages} {052212} (\bibinfo {year} {2019})}\BibitemShut {NoStop}%
\bibitem [{\citenamefont {Marvian}(2022)}]{marvian2022restrictions}%
  \BibitemOpen
  \bibfield  {author} {\bibinfo {author} {\bibfnamefont {I.}~\bibnamefont
  {Marvian}},\ }\bibfield  {title} {\bibinfo {title} {Restrictions on
  realizable unitary operations imposed by symmetry and locality},\ }\href
  {https://doi.org/10.1038/s41567-021-01464-0} {\bibfield  {journal} {\bibinfo
  {journal} {Nature Physics}\ }\textbf {\bibinfo {volume} {18}},\ \bibinfo
  {pages} {283} (\bibinfo {year} {2022})}\BibitemShut {NoStop}%
\bibitem [{\citenamefont {Cheng}\ and\ \citenamefont
  {Swingle}(2021)}]{cheng2021scrambling}%
  \BibitemOpen
  \bibfield  {author} {\bibinfo {author} {\bibfnamefont {G.}~\bibnamefont
  {Cheng}}\ and\ \bibinfo {author} {\bibfnamefont {B.}~\bibnamefont
  {Swingle}},\ }\bibfield  {title} {\bibinfo {title} {Scrambling with
  conservation laws},\ }\href {https://doi.org/10.1007/JHEP11(2021)174}
  {\bibfield  {journal} {\bibinfo  {journal} {J. High Energy Phys.}\ }\textbf
  {\bibinfo {volume} {2021}}\bibinfo  {number} { (11)},\ \bibinfo {pages}
  {1}}\BibitemShut {NoStop}%
\bibitem [{\citenamefont {Alex}\ \emph {et~al.}(2011)\citenamefont {Alex},
  \citenamefont {Kalus}, \citenamefont {Huckleberry},\ and\ \citenamefont {von
  Delft}}]{Alex_2011}%
  \BibitemOpen
\bibfield  {number} {  }\bibfield  {author} {\bibinfo {author} {\bibfnamefont
  {A.}~\bibnamefont {Alex}}, \bibinfo {author} {\bibfnamefont {M.}~\bibnamefont
  {Kalus}}, \bibinfo {author} {\bibfnamefont {A.}~\bibnamefont {Huckleberry}},\
  and\ \bibinfo {author} {\bibfnamefont {J.}~\bibnamefont {von Delft}},\
  }\bibfield  {title} {\bibinfo {title} {A numerical algorithm for the explicit
  calculation of su(n) and sl(n,c) clebsch–gordan coefficients},\ }\href
  {http://dx.doi.org/10.1063/1.3521562} {\bibfield  {journal} {\bibinfo
  {journal} {Journal of Mathematical Physics}\ }\textbf {\bibinfo {volume}
  {52}},\ \bibinfo {pages} {023507} (\bibinfo {year} {2011})}\BibitemShut
  {NoStop}%
\bibitem [{\citenamefont {Chen}\ and\ \citenamefont
  {Zhou}(2019)}]{chen2019quantum}%
  \BibitemOpen
  \bibfield  {author} {\bibinfo {author} {\bibfnamefont {X.}~\bibnamefont
  {Chen}}\ and\ \bibinfo {author} {\bibfnamefont {T.}~\bibnamefont {Zhou}},\
  }\bibfield  {title} {\bibinfo {title} {Quantum chaos dynamics in long-range
  power law interaction systems},\ }\href
  {https://link.aps.org/doi/10.1103/PhysRevB.100.064305} {\bibfield  {journal}
  {\bibinfo  {journal} {Phys. Rev. B}\ }\textbf {\bibinfo {volume} {100}},\
  \bibinfo {pages} {064305} (\bibinfo {year} {2019})}\BibitemShut {NoStop}%
\end{thebibliography}%
\onecolumngrid
\newpage
\appendix
\begin{appendices}

\section{Anti-commuting fermions on multiple time-contours}
\label{App:four_time_fermions}
This short section will elucidate the procedure to arrive at anti-commuting fermions on four time contours. In the original model, we begin with a Brownian model built with fermions $\chi_{i,r}$ where $i$ is the fermion index and $r$ is the site/cluster index. These satisfy the anti-commutation relation $\{ \chi^{\dagger}_{i,r},\chi_{j,r'} \} = \delta_{i,j} \delta_{r, r'}$. When generalizing to four time contours, we use the 'replica fermions' on replicas $a,b,c,d$:
\bea
&\chi_{i,r}^{a}:=\chi_{i,r} \otimes I \otimes I \otimes I \qquad  \chi_{i,r}^{b}:=I \otimes \chi_{i,r}^{\top} \otimes I \otimes I \\
&\chi_{i,r}^{c}:=I \otimes I \otimes \chi_{i,r} \otimes I \qquad  \chi_{i,r}^{d}:=I \otimes I \otimes I \otimes \chi_{i,r}^{\top}
\eea
where $\chi^{\top}$ refers to the transpose, and we have performed a particle-hole transformation on replicas $b,d$ for future mathematical convenience. It's clear that the fermions on different replicas commute with each other due to the tensor-product structure. To make them anti-commute, we use the global parity operator $\Q$
\bea
\Q^{\alpha} = \prod_{j,r} \text{exp}(i \pi n^{\alpha}_{j,r}), \quad \alpha = a,b,c,d
\eea
Following this, we define the fermions $\psi^{\alpha}_{i,r}$ using the parity operator:
\bea
&\psi_{j}^{a}=\mathcal{Q}^{a} \chi_{j}^{a}, \quad  &&\psi_{j}^{b}=\mathcal{Q}^{a} \chi_{j}^{b}, \\
&\psi_{j}^{c}=\mathcal{Q}^{a} \mathcal{Q}^{b} \mathcal{Q}^{c} \chi_{j}^{c}, \quad &&\psi_{j}^{d}=\mathcal{Q}^{a} \mathcal{Q}^{b} \mathcal{Q}^{c} \chi_{j}^{d}.
\eea
These new fermions now anti-commute even on separate replicas,
$\{ \psi^{\dagger \alpha}_{i,r},\psi_{j,r'}^{\beta} \} = \delta_{i,j} \delta_{\alpha, \beta} \delta_{r, r'}$, and can be used to define the SU(4) algebra as highlighted in the main text.

\section{Charge-resolution of null-eigenstates}
\label{App:Identity}
In this section of the appendix, we will discuss the properties of the Identity operator state and other null-eigenstates in a model with charge conservation. As was noted in Sec.~\ref{sec:operator_basis}, we work with an operator basis with a well defined left and right charge, which are both constants of motion. Since the identity is invariant due to the unitary dynamics (Fig.~\ref{Fig:OTOC-Tensor}), the corresponding operator state vanishes under the action of the emergent Hamiltonian:
\bea
U I U^{\dagger} &= I \implies \overline{U \otimes U^*} \ket{I } = \ket{I} \implies \H \ket{I } = 0 \, ; \,
\ket{I} &= \sum_{i,j} \delta_{ij} \ket{i \otimes j}
\eea
While the above statement is true for any unitary Brownian model, in a charge-conserving model this identity splits into multiple charge sectors. For a complex fermionic model with $N$ fermions:
\bea
I = \prod_{i=1}^{N} (n_i + \bn_i) = \sum_{q=0}^N I_{q,N} \; ; \; I_{q,N} = \sum_{i_1<...<i_q} n_{i_1}...n_{i_q}\bn_{i_{q+1}}...\bn_{i_{N-q}}
\eea
where we have split the identity into strings $I_{q,N}$, each with charge profile $(q,q)$. Since both the identity and the charge profiles are conserved, each string $I_{q,N}$ is also a constant of motion $\H \ket{I_{q,N}} = 0$. This means that due to charge conservation, we have multiple constants of motion from the invariance of the identity, instead of just one. There exists one in each sector with profile $(q,q)$, hence $q_a = q_b$ is a necessary condition to restrict to a sector with a component of the identity. Apart from the identity, any other operator which represents a symmetry will also have corresponding charge-resolved null eigenstates. On the four-contour level, there emerges another null-eigenstate of the emergent Hamiltonian which is due to the unitary nature of the dynamics as well. This is the complete set of states $\sum_{\S} \ket{\S^{\dagger} \otimes \S}$ (Fig.~\ref{Fig:OTOC-Tensor}), and this as well will split into different charge sectors, each with profile of the form $(q_1, q_2, q_2, q_1)$. This is because the charge profiles of an operator and its complex conjugate are related in the following way:
\bea
(\sum_i n_i) O = q_a O \implies O^{\dagger} (\sum_i n_i) = q_a O^{\dagger}
\eea

\section{Solving the restricted Fokker-Planck equation}
\label{App:Restr_FP}

In this section, we will study simple toy versions of the 'restricted' differential equations we obtain in the main section. The goal is to show that once we restrict to the sector which describes a stochastic process, the usual process of obtaining a Langevin equation still remain valid.

Let's consider a function $P(x,y,t)$ which satisfies the following differential equation:
\bea
\partial_t P = \partial_x ( (xy +x) P) + \partial_y ( y P)
\eea
Hence $P$ describes a probability distribution since $\int P dx dy$ is a constant. It is well known that the equations of motion for $x,y$ can be described by the followin Langevin equations:
\bea \label{eq:Langevin}
x'(t) = - (x y + x) ; \quad y'(t) = -y
\eea
Now we modify the equation such that it only describes a probability distribution in the sector $y=0$, which means it is the incoherent sector:
\bea
\partial_t P = \partial_x ( (xy +x) P) + y \partial_y P \implies \partial_t \int \delta(y) P dx dy = 0
\eea
We begin with an input state which is well localized around the fixed probability sector $P(x,y,t=0) \propto e^{-y^2} \delta(x-\epsilon)$, and we are interested in computing the quantity $\F$ is given by $\F(t) = \int \delta(y) x P(x,y,t) dx dy$. This model is a simplified version of the situation in the main text, with $y=0$ representing the `incoherent' sector. The restricted differential equation above supports solutions of the form:
\bea
P(x,y,t) = \frac{1}{x} f \bigg(y - \text{ln}\bigg(\frac{x}{y}\bigg) , \, \text{ln} \, y +t \bigg) = \frac{1}{x} \delta \bigg( \text{ln} \, x + y(e^t-1) + t - \text{ln} \, \epsilon \bigg) e^{-y^2 e^{2t}}
\eea
where we have reshaped $P(x,y,t=0)$ to be of the from $f$, time-evolved it, and then simplified it to be the expression on the right. Hence $\F$ is given by:
\bea
\F (t) = \int x \delta(y) P(x,y,t) dx dy= \int x \, \delta(x - e^{-t} - \epsilon) dx
\eea
which implies that $x$ satisfies the equation of motion $x'(t) = -x$ which is precisely what one gets upon setting $y=0$ in Eq.~\eqref{eq:Langevin}. The overall equations of motion have changed, but in the fixed probability sector, variables follow Langevin equations as expected, and there is no `feedback' effect from the non-conserved sector.

\section{Analytic solution of the free-fermionic chain}
In this section we analytically solve the transport properties and OTOC in the free fermionic chain which is represented by the Hamiltonian:
\bea
H_{\text{inter}} &= \sum_{j_1,j_2,r} K^{r}_{j_1,j_2}(t) \chi^{\dagger}_{j_1,r}\chi_{j_2,r+1} + h.c.
\eea
This represents a fermionic model with $N$ fermions on each cluster, while driven hopping is allowed between $L$ such clusters.

\subsection{Charge transport at finite-$N$}
The transport occurs on two-time contours, as it can be diagnosed by analyzing the two-point function of the charge density:
\bea
\langle \rho(r_1,t) \rho(r_2,0) \rangle = \frac{\tr(\rho(r_1,t) \rho(r_2,0))}{\tr(I)} ;\quad \rho = \sum_{i=1}^{N} \frac{n_i}{N}
\eea
For operator dynamics in the Brownian model, we consider the two-replica level emergent Hamiltonian, which will show an emergent SU(2) $\otimes$ U(1) symmetry. Defining the operator $S^{\alpha \beta} = \sum_{i} {\chi^{\dagger}_i}^{\alpha} \chi_i^{\beta}$, we can show that it can be used to define the algebra as follows:
\bea
L^{+} = S^{ab} ;\quad  L^{-} = S^{ba}; \quad [L^{+},L^{-}]=2L_{z}=S^{ab}-S^{ba}; \quad Q = S^{ab}+S^{ba}
\eea

where $Q$ is the U(1) part of the algebra. In the SU(2) algebra on each site, the input state $\ket{\rho}$ can be decomposed as:
\bea
&\ket{n_1} = \ket{n_1(n_2+\bn_2)...(n_N+\bn_N)} = \frac{1}{\sqrt{2}} \ket{\uparrow \rightarrow ... \rightarrow}\\
&=\frac{1}{2} \ket{(\rightarrow + \leftarrow) \rightarrow ... \rightarrow}= \frac{1}{2} \ket{\rightarrow \rightarrow ... \rightarrow} + \frac{1}{2}\ket{\leftarrow \rightarrow ... \rightarrow}\\
&\ket{\leftarrow \rightarrow ... \rightarrow} = \frac{1}{N} \sum_{i} \ket{\rightarrow...\leftarrow_i...\rightarrow} + \bigg(\ket{\leftarrow \rightarrow...\rightarrow}- \frac{1}{N} \ket{\rightarrow...\leftarrow_i...\rightarrow}\bigg)\\
&\implies \ket{\rho} = \frac{1}{2} \ket{N/2,N/2}_x + \frac{1}{2 \sqrt{N}} \ket{N/2,N/2-1}_x\\
&\implies \ket{\rho(r_1)} = \ket{I}_1 \otimes \ket{I}_2 \otimes...\ket{\rho}_{r_1}...\otimes \ket{I}_L
\eea
The identity is the fully polarized state along the $x$-direction on every site, and is a constant of motion for the unitary dynamics (Appendix \ref{App:Identity}). The trace is given by:
\bea
\frac{\tr(\rho(r_1,t) \rho(r_2,0))}{\tr(I)} = \langle \rho(r_2) | U \otimes U^* | \rho(r_1) \rangle
\eea
One can verify that these give the correct values at $t=0$ according to the spin decomposition, i.e. 1/4 for $r_1 \neq r_2$ and 1/4(1+1/N) for $r_1=r_2$. Now that we have the state-decomposition, we can analyze the emergent Hamiltonian, which for the free fermionic chain is:
\bea
\H =  \frac{1}{N}\bigg(\sum_{r=1}^{L} 2L^z_r L^z_{r+1} + L^{+}_r L^-_{r+1} + L^{-}_r L^+_{r+1} - \frac{N^2 L}{2}\bigg)
\eea
This is simply the SU(2) Heisenberg model, and we can exploit the SU(2) invariance to solve the two-point function exactly. The 'physical' charge conservation manifests itself as conservation of global weights $\sum_r L_r^i=$ const, $i=x,y,z$. This should be contrasted with the mathematical charge conservation ($Q$) within each irrep. Since the input state of interest splits into two states, one of which is static under the emergent Hamiltonian, we only need to focus on the lower weight state. We define a basis:
\bea
\ket{\psi_r} = \ket{N/2}_{x,1} \otimes...\ket{N/2-1}_{x,r} \otimes...\ket{N/2}_{x,L}
\eea
such that a general state may be written as: 
\bea
\ket{\psi(t)} = \sum_{r} q_r(t) \ket{\psi_r} \implies \sum_r q_r(t) = const,
\eea
hence $q_r(t)$ can be viewed as the charge distribution. The equation of motion for the state is:
\bea
&\partial_t \ket{\psi(t)} = \H \ket{\psi(t)} \implies \sum_r \partial_t q_r(t) \ket{\psi_r} =  \sum_i q_r(t) ( \ket{\psi_{r+1}} +  \ket{\psi_{r-1}} -2 \ket{\psi_{r}})\\
&\implies \partial_t q_r(t) = (q_{r+1}(t)+q_{r-1}(t)-2q_r(t)) \longrightarrow \partial_t \rho(r,t) =  \partial^2_r \rho(r,t)
\eea
where the last equation involves taking the continuum limit of the spatial direction $r$. Starting with a delta function localized at $r=r_1$, we get the solution:
\bea
\rho(r,t) = \frac{1}{\sqrt{8 \pi N t}} \, \text{exp}\bigg(-\frac{(r-r_1)^2}{4 t}\bigg) 
\eea
which means the expression for the 2-pt func. is:
\bea
\langle \rho(r_1,t) \rho(r_2,0) \rangle = \frac{1}{4}+\frac{1}{4N\sqrt{4 \pi t}} \text{exp}\bigg(-\frac{(r_2-r_1)^2}{4 t}\bigg) 
\eea

\subsection{OTOC at finite-$N$}
\label{App:free_fermion}
In this section we will exactly solve the OTOC for the free fermion at finite $N$, for a chain with $L$ clusters. We will choose both the input operator and output operator, $W$ and $V$, to be $\chi$ but on clusters $r$ and $r'$, because it makes the computations simpler. The Emergent Hamiltonian for the free-fermion case, assuming periodic boundary conditions, is given by:
\bea
\H_{\text{inter}}= \frac{1}{N} \bigg( \sum_{\alpha,\beta,r} S^{\alpha \beta}_r S^{\beta \alpha}_{r+1}-N \sum_r Q_r \bigg)
\eea
This has been written explicitly in terms of the Casimir of SU(4) to make the SU(4) invariance manifest. The OTOC we will study will be of the form $\sum_{i,j} \F(\chi_{i,r}, \chi_{j,r'})$. The input state is the operator-state $\ket{\chi^{\dagger} \otimes \chi}$ on site $r$ and the identity $\ket{I \otimes I}$ on all the other sites.
\bea
\ket{\i_{r}} &= \frac{1}{2^{NL}} \ket{I \otimes I}_1 \otimes \ldots  \ket{\chi^{\dagger} \otimes \chi}_{r}  \otimes \ldots \ket{I \otimes I}_L ; \\
\ket{\chi^{\dagger} \otimes \chi}_{r} &= \frac{1}{N} \sum_{i=1}^{N} \ket{\chi_i^{\dagger} \otimes \chi_i}_{r}.
\eea
The second line is due to the restriction of the operator to the symmetric irrep $(0,N,0)$. Note that since the identity-state is automatically contained within this irrep, we have fixed the irrep to be the same on every site. This overall state is complicated in terms of its decomposition into different charge sectors which have independent dynamics. Now we utilize the SU(4) invariance of the Emergent Hamiltonian and specifically the SU(2) $\otimes$ SU(2) sub-algebra formed by the operators $S^{ba}$ and $S^{dc}$ (Fig.~\ref{Fig:Complex-Spin}) to rotate the input state into one strict charge-sector:
\bea
\ket{\widetilde{\i}_{r}} = \frac{1}{2^{NL}} \ket{\Pi n \otimes \Pi n}_1 \otimes \ldots \ket{\chi^{\dagger} \Pi n \otimes \chi \Pi n}_{r} \otimes \ldots \ket{\Pi n \otimes \Pi n}_L
\eea
where $\ket{\widetilde{\i}}_{r}$ is the rotated input state. We have simply rotated the identity $I = n + \bn$ into $n$ on both the left and right halves of the string via the SU(2) $\otimes$ SU(2) rotation. This leaves the emergent Hamiltonian unchanged due to the SU(4) invariance. We will now demonstrate that this state performs a random walk under the dynamics. Let there be a general state given by:
\bea
\ket{\widetilde{\psi} (t)} = \sum_r c_r(t) \ket{\widetilde{\i}}_{r}\\
\partial_t \ket{\widetilde{\psi} (t)} = \H \ket{\widetilde{\psi} (t)}
\eea
where on the second line we have demonstrated the rule which governs the dynamics of the said state. Since the input-state now has fixed values of $\sum_r S_r^{\alpha \alpha} S_{r+1}^{\alpha \alpha}$ after the rotation, it is simple to check that it satisfies:
\bea \label{eq:diffusive_OTOC}
 &\H (\sum_r c_r(t) \ket{\widetilde{\i}_{r}}) =  \sum_r c_r(t) (\ket{\widetilde{\i}_{r+1}} +\ket{\widetilde{\i}_{r-1}}-2 \ket{\widetilde{\i}_{r}})\\
 &\implies \partial_t c_r(t) = (c_{r+1} + c_{r-1} - 2 c_r) \rightarrow \partial_t c(r,t) = \partial_r^2 c(r,t)
\eea
In the second line we have used the continuum limit along the spatial dimension $r$. This result is exact for all $N$ and $L$. One can now rotate the states back into the form $\ket{\i}_r$ and show that it performs a random walk under the time evolution. Since the output operator is also chosen to be $\chi$, it is simple to verify that it satisfies the following overlap condition:
\bea
 2^{NL} \braket{\o_{r'}|\i_{r}}= 
\begin{cases}
    \frac{1}{4N}-\frac{1}{4},& \text{if } r=r'\\
    -\frac{1}{4},& \text{otherwise} 
\end{cases}
\eea
Hence the overall overlap with the state $\psi$, which is the state $\widetilde{\psi}$ rotated back, takes the form:
\bea
\F(W(t),V) = 2^{NL} \braket{\o_{r'}|\psi(t)} \\ = -\frac{1}{4} \sum_r c_r(t) + \frac{1}{4N} c_{r'}(t)
\eea
The sum of the coefficients $c_r$ is conserved due to the random-walk nature of the dynamics and the initial state can be chosen to guarantee that it is fixed at 1. Hence the OTOC in the continuum limit is determined by the value of $c(r',t)$ which shows diffusive growth/decay (Eq.~\eqref{eq:diffusive_OTOC}). The exact OTOC at finite $N$ therefore matches the OTOC in the infinite $N$ limit.

\end{appendices}
\end{document}